\documentclass[12pt,preprint2,a4paper,trackchanges]{aastex6}
\usepackage{graphics}
\usepackage{amsmath, amsthm, amssymb}
\usepackage{epsfig}

\usepackage{graphicx}

\begin{document}

\title{Integrated spectra of Milky Way globular clusters}

\author{T. C. Moura\altaffilmark{1}}
\affil{Universidade de S\~ao Paulo, IAG, Rua do Mat\~ao 1226,
Cidade Universit\'aria, S\~ao Paulo 05508-900, Brazil}

\author{M. Trevisan\altaffilmark{2}}
\affil{Universidade Federal do Rio Grande do Sul, 
Departamento de Astronomia, CP 15051, Porto Alegre 91501-970, Brazil}

\author{B. Barbuy\altaffilmark{1}, S. Rossi\altaffilmark{1}}
\affil{Universidade de S\~ao Paulo, IAG, Rua do Mat\~ao 1226,
Cidade Universit\'aria, S\~ao Paulo 05508-900, Brazil}

\received{2019 July 16}
\revised{2019 August 20}
\accepted{2019 August 26}
\published{2019 October 25}

\begin{abstract}
Integrated spectra of Milky Way globular clusters (GCs) are reproduced
by computing synthetic spectra taking into account individual element
abundances. Five
clusters were selected from their location in the Galactic bulge, for which
integrated spectra were available in the WiFeS Atlas of Galactic
Globular Cluster Spectra project. Our aim is to further study the oldest
GC located in the Galactic bulge, with a metallicity in the
range $-$1.6$<$[Fe/H]$<$$-$0.7. We also include the halo cluster NGC 6752
for comparison purposes.
We reproduce the full spectra in the range 4500-9000 {\rm \AA} available
in these observed spectra, as well as individual lines of Na, Mg, Al, Si, Ca,
Ti, Ba, and Eu. We report a list of lines that are suitable for
abundance derivation, and by adopting these abundances we are able to
fit the damping parameters that define the wings of strong lines of
well-known triplets of Mg I and Ca II. Finally, the effect of multiple stellar
populations through enhanced Na abundances is tested.
\end{abstract}


\section{Introduction}

Integrated spectra of Milky Way (MW) globular clusters (GCs) are becoming an increasingly
important ingredient in the study of faint and compact
GCs in our Galaxy and in other galaxies, as well as the integrated spectra of galaxies.

The integrated spectra of MW GCs observed by Bica \& Alloin (1986), Schiavon et al. (2005), and
Usher et al. (2017) offer an important basis for the validation of
codes that try to reproduce integrated spectra of star clusters and
old galaxies.

At the same time, given the degeneracy between
  metallicity and age, in order to reproduce the integrated spectra, 
it is best to have input data on ages and chemical abundances,
which is possible for MW and Magellanic Cloud GCs.
Compilations of such data for the MW
were presented by Pritzl et al. (2005) and
Harris (1996, edition of
  2010)\footnote{www.physics.mcmaster.ca/$\sim$harris/mwgc.dat}.
 In a series of papers
Saviane et al. (2012), Dias et al. (2015, 2016)
and V\'asquez et al. (2018) derived abundances and alpha-to-iron ratios
from FORS2@VLT spectra for individual stars in 61 GCs,
targeting less well-studied ones, and compared them with the literature.

Work on derivation of abundances from integrated spectra is becoming
  available for the old GCs in the Galaxy, where ages
  from the literature are adopted.
Roediger et al. (2014) derived abundances from the integrated
  spectra of 41 GCs for which Schiavon et al. (2005)
had obtained the integrated spectra. Sakari et al. (2013) and Colucci et al.
(2017, and references therein) were able to measure equivalent widths (EW)
and derive abundances. Larsen et al. (2017, and references therein)
and Conroy et al. (2018, and references therein) were able to
derive metallicities and abundances from full spectrum fitting of
integrated spectra of GCs.

A recent review on the building of integrated spectra for single stellar
populations (SSPs) is presented by Martins et al. (2019). They
investigate the methods and ingredients used, including
differences between  evolutionary tracks or
use of observed colour-magnitude diagrams (CMDs), initial mass functions (IMFs)
and spectral libraries available in the literature.

In the present work we use spectra from
the WiFeS Atlas of Galactic Globular 
Cluster Spectra (WAGGS) project by Usher et al. (2017), that made available
integrated spectra of a series of GCs. We have selected
for our analysis, six GCs from Usher et al. (2017)
that have good S/N ratios, where five of them are contained in the
list of 43 bulge GCs as selected by Bica et al. (2016),
and additionally the halo cluster NGC 6752, for comparisons with the
literature. 
We also limited the metallicity to be within $-$1.6$<$[Fe/H]$<$$-$0.7, in order
to concentrate efforts on the oldest GCs of the Galactic bulge.

For the building of the integrated spectra, 
we here describe in detail the code SynSSP.
 In this code we adopt the Dartmouth
(Dotter et al. 2008) isochrones, and compute the spectra of a series
of stars along the main sequence (MS), subgiant ( and red giant phases, for
a given age, metallicity, [$\alpha$/Fe] and element abundances.
The synthetic spectra are computed in each run.
A previous version of this procedure was already described in
La Barbera et al. (2013). The code is now much improved,
  mainly due to the  update of the code PFANT for
  spectrum synthesis, as described in Barbuy et al. (2018b).
  The code was entirely upgraded with homogeneization of language to
  Fortran 2003, optimization for speed, improved error reporting
  in case of problems with the input data files. Line lists were also
  revised.
  
The ages are adopted from the recent age derivations by
Kerber et al. (2018), R. A. P. Oliveira et al. (2019, in preparation), and VandenBerg et al. (2013),
from isochrone fitting to CMDs.
Our main interest is to reproduce the integrated spectra by computing 
stellar spectra taking into account their elemental abundances,
reported in high-resolution spectroscopic studies, and by further deriving
element abundances from the integrated spectra.

We also investigate the multiple stellar populations, in proportions
derived by Milone et al. (2017), and R. A. P. Oliveira et al. (2019, in preparation), by analyzing
the effect of a combined spectrum of first and second generations, with
different Na abundances. The use of Na and TiO line intensities to
deduce a bottom-heavy stellar IMF is also discussed.

In Section $\,$2 the available literature on the sample clusters is reported. In Section 3 we
describe the calculations of integrated spectra of simple stellar populations employed in
this work. In Section 4 we examine individual abundances that can be derived from the integrated
spectra. In Section 5 the effect of multiple stellar populations on Na lines
is investigated. In Section 6 conclusions are drawn.

\section{Sample globular clusters}

Our sample is composed of five GCs in the Galactic bulge and
one halo cluster in the
metallicity range $-$1.6$\lesssim$ [Fe/H] $\lesssim$$-$0.7, with
ages older than 11 Gyr, and for which there are available integrated
spectra in the
WiFeS Atlas of Galactic Globular  Cluster Spectra - Waggs project from Usher
et al. (2017).
The integrated spectra from the WAGGS library were observed with the WiFeS integral field spectrograph installed in the Australian National University 2.3 m telescope and cover a wide wavelength range from 3300 - 9050 {\rm \AA}{} at a spectral resolution of
R$\sim$6,800. 

Basic information on the selected GCs, including the galactic coordinates,
distances to the Sun and to the Galactic centre,
and absolute magnitudes M$_{\rm V}$ from 
Harris (1996) are given in the Table \ref{tab1}.
In this Table are also given the fractions of
first generation (1G) stellar populations relative to total,
as estimated by Milone et al. (2017)
from {\it Hubble Space Telescope} ultraviolet filters, and ages deduced from
Kerber et al. (2018), R. A. P. Oliveira et al. (2019, in preparation), and VandenBerg et al. (2013).
Finally, He enhancements from a second generation (2G) relative to 1G
from Milone et al. (2019) are given.
Literature chemical abundances derived with different methods 
are reported in Table \ref{tab2}.

\begin{table*}[ht!]
 \begin{center}
  \caption{Galactic coordinates, Distances to the Sun and to the Galactic Center, Absolute Visual Magnitude  from Harris (1996), Fraction of First Generation Stars from Milone et al. (2017), Ages from R. A.P. Oliveira et al. (2019, in preparation) for NGC 6624, NGC 6637 and NGC 6723, Ages from Kerber et al. (2018) for NGC 6522, and  Ages from VandenBerg et al. (2013) for NGC 6171 and NGC 6752}
  \label{tab1}
  \setlength{\tabcolsep}{10pt} 
  \begin{tabular}{lccccccccc}
  \tableline
  \hline
   Cluster     & l($\circ$) & b($\circ$) & \hbox{d$_{\rm \odot}$} & \hbox{d$_{\rm GC}$} &  
   M$_{\rm V}$ & {\rm N$_1$/N$_{TOT}$} & Age & $\delta$Y$_{1G,2G}$ \\
               &            &            & (kpc)                  & (kpc)               &
               &                       & (Gyr) & \\
  \hline
   NGC 6171 & 3.37   & 23.01   & 6.4  & 3.3 &  -7.12 & 0.397 & 12.0  & 0.019 \\
   NGC 6522 & 1.02   & -3.93   & 7.7  & 0.6 &  -7.95 & 0.160 & 13.0  & --- \\
   NGC 6624 & 2.79   & -7.91   & 7.9  & 1.2 &  -7.49 & 0.279 & 12.5  & 0.010 \\
   NGC 6637 & 1.72   & -10.27  & 8.8  & 1.7 &  -7.64 & 0.425 & 12.4  & 0.004 \\ 
   NGC 6723 & 0.07   & -17.30  & 8.7  & 2.6 &  -7.83 & 0.363 & 12.5  & 0.005 \\
   NGC 6752 & 336.49 & -25.63  & 4.0  & 5.2 &  -7.73 & 0.294 & 12.5  & 0.010,0.032 \\
 \tableline
 \multicolumn{9}{l}{Note. Y are helium enhancements in second generations stars (and third for the case of NGC 6752) from Milone et al. (2018, 2019).}
  \end{tabular}
 \end{center}
\end{table*}

\begin{table*}
\begin{center}
\centering
  \caption{Literature Metallicities and Abundances for the Sample Clusters}
  \label{tab2}
    \setlength{\tabcolsep}{2pt} 
 \begin{tabular}{ccccccccccccc}
 \tableline
 \hline
 Cluster  & [Fe/H] & [O/Fe] & [Na/Fe] & [Al/Fe] & [Mg/Fe] & [Si/Fe] & [Ca/Fe] & [Ti/Fe] &  [Ba/Fe] & [Eu/Fe] & method & reference \\
 \hline
 NGC 6171 & -1.02  & +0.17    & +0.37    & ---    & +0.51    & +0.54    & +0.06    & +0.40   & ---    & ---  & comp. &   Roediger et al. (2014) \\ 
& -1.02   & --- & --- & ---  & +0.41 & +0.25 & +0.32   & +0.42 & --- & ---    & ISFF & Conroy et al. (2018) \\
         & -1.58..  & ---    & --- & ---  & ---  & --- & +0.24  & --- & --- & ---  & ISEW & Usher et al. (2019) \\ 
NGC 6522 & -0.95  & +0.36    & +0.05     & +0.20   & +0.23    & +0.13    & +0.13    & +0.04   & +0.32   & +0.30  & HRS & Barbuy et al. (2014) \\  
         & -1.12  & ---    & +0.24     & +0.62  & +0.42    & +0.28    & +0.35   & +0.38  & +0.37  & +0.43  & HRS & Ness et al. (2014) \\  
(2G)     & -1.04  & +0.33  & ---    & +0.57  & +0.13   & +0.30   & ---    & ---  & ---   & ---  & HRS & Fernandez-Trincado et al. (2018) \\  
& -1.21   &  ---   & ---     & ---     & +0.47 & +0.30 & +0.35   & +0.40 & --- & --- & ISFF & Conroy et al. (2018) \\
         & -1.58..  & ---    & --- & ---  & ---  & --- & +0.13  & --- & --- & ---  & ISEW & Usher et al. (2019) \\ 
         & -1.34  & +0.49  & +0.04 & --- & +0.27  & +0.25  & +0.17 & +0.16   & --- & --- & comp.& Roediger et al. (2014) \\    
NGC 6624 & -0.69  & +0.41    & $-$      & +0.39   & +0.42    & +0.38    & +0.40    & +0.37   & $-$    & $-$  & HRS  & Valenti et al. (2011)  \\  
& -0.77    & ---    &  ---    & --- & +0.36 & +0.25 & +0.23  & +0.41 & --- & ---    & ISFF & Conroy et al. (2018) \\
         & -1.58..  & ---    & --- & ---  & ---  & --- & +0.18  & --- & --- & ---  & ISEW & Usher et al. (2019) \\ 
         & -0.44  & +0.41    & --- & --- & +0.42 & +0.38 & +0.40 & +0.37   & --- & --- & comp.& Roediger et al. (2014) \\  
NGC 6637 & -0.77  & +0.20    & +0.35    & +0.49   & +0.28    & +0.45    & +0.20    & +0.24   & +0.22  & +0.45  & HRS & Lee et al. (2007) \\      
& -0.90  & ---    & --- & --- & +0.43 & +0.20 & +0.30 & +0.54 & --- & --- & ISFF & Conroy et al. (2018) \\
         & -1.58..  & ---    & --- & ---  & ---  & --- & +0.20  & --- & --- & ---  & ISEW & Usher et al. (2019) \\ 
      & -0.64  & +0.20  & +0.35 & --- & +0.28 & +0.45 & +0.20 & ---   & ---    & --- & comp.& Roediger et al. (2014) \\  
NGC 6723 & -0.98  & +0.29    & +0.00     & 0.31    & +0.23    & +0.36    & +0.30    & +0.24     & +0.22  & $-$    & HRS & Rojas-Arriagada et al. (2016) \\  
& -1.22  & +0.39    & +0.05     & ---    & +0.58    & ---    & ---   & ---    & ---  & ---   & HRS & Gratton et al. (2015) (BHB) \\ 
& -1.22  & +0.55    & +0.11     & ---    & +0.50    & +0.59  & +0.81  & ---    & +0.75  & ---   & HRS & Gratton et al. (2015) (RHB) \\
(1G)& -0.91  & +0.39    & +0.14     & +0.32    & +0.47    & +0.52  & +0.37  & +0.34   & +0.36  & +0.38   & HRS & Crestani et al. (2019) \\   
& -1.32   & ---    & --- & --- & +0.53    & +0.20 & +0.33  & +0.29 & --- & --- & ISFF & Conroy et al. (2018) \\
         & -1.58..  & ---    & --- & ---  & ---  & --- & +0.49  & --- & --- & ---  & ISEW & Usher et al. (2019) \\ 
       & -1.10   & ---    & ---    & ---  & +0.44   & +0.68  & +0.33 & +0.24   & --- & --- & comp.& Roediger et al. (2014) \\  
NGC 6752 & -1.56  & +0.16  & +0.33   & +0.41 & +0.50  & +0.38    & ---    & ---   & --- & --- &HRS& Carreta et al. (2009) \\
     (1G) & -1.56  & +0.54  & +0.01   & +0.11 & +0.45  & +0.51    & ---    & ---   & --- & --- &HRS& Carretta et al. (2012) \\
    (2G) & -1.56  & +0.25  & +0.36   & +0.74 & +0.42  & +0.48    & ---    & ---   & --- & --- &HRS& Carretta et al. (2012) \\
    (3G) & -1.56  & +0.06  & +0.62   & +1.21 & +0.31  & +0.43    & ---    & ---   & --- & --- &HRS& Carretta et al. (2012) \\
& -1.58  & ---    & +0.09 & ---  & ---  & --- & +0.46  & +0.36 & +0.10 & ---  & ISEW & Colucci et al. (2017) \\
         & -1.58..  & ---    & --- & ---  & ---  & --- & +0.26  & --- & --- & ---  & ISEW & Usher et al. (2019) \\ 
         & -1.883  & ---   & +0.302 & ---  & +0.391  & --- & +0.355 & +0.342   & +0.182 & --- & ISFF & Larsen et al. (2017) \\    
         & -1.64 & --- & --- &  --- & +0.54     & +0.42    & +0.33 & +0.29 & --- &  --- & ISFF & Conroy et al. (2018) \\  
  \tableline
\end{tabular}
\end{center}
\end{table*}

\section{Integrated spectrum model}

In order to reproduce the integrated spectrum of a stellar population, the following ingredients are required: stellar evolutionary tracks and a library of stellar spectra. The isochrones determine the mass and the atmospheric parameters (surface gravity, $\log g$, and effective temperatures, $T_{\rm eff}$) of stars belonging to a SSP with a given age, metallicity, and [$\alpha$/Fe] enhancement.
A number of stellar evolutionary stages are defined
to be included in the synthesis, such that the integrated light spectrum of the SSP
can be generated by combining the spectra of individual stars, 
according to a given IMF. In the present work
we assume a Salpeter IMF (Salpeter 1955) which is very similar with Kroupa (2001) in the mass range of old stars.

Spectra of individual stars from empirical or synthetic libraries can be used. Given our aim of fitting particular 
lines by varying their abundances, 
empirical stellar libraries are not well suited for investigating how changes in the abundances of individual elements affect the spectral indices, since they are restricted to abundance patterns of stars in the solar neighborhood. 

For this reason, we developed the {\sc SynSSP} package, which computes the spectra of SSPs with variable abundance ratios. 
The user can compute a synthetic SSP spectra by providing some key parameters, as follows:

A. Stellar evolutionary ingredients:

\begin{itemize}
 \item Age, metallicity and [$\alpha$/Fe] of the SSP. 
 We adopt the Dartmouth\footnote{\url{http://stellar.dartmouth.edu/models/index.html}} stellar evolutionary tracks (Dotter et al. 2008).
  
\item Number of stars, characterized by their effective temperatures $T_{\rm eff}$, and surface gravity $\log g$, that should be selected along the isochrone. A standard procedure considers 12 stars in the main sequence (MS) and subgiant branch (SGB), and 9 stars in the red giant branch (RGB).

\end{itemize}

 B. Synthetic stellar spectra: 
 
 \begin{itemize}
 
 \item A stellar synthetic spectrum is  computed for each of these $T_{\rm eff}$, $\log g$ pairs. 
 
 \item The values of [C, N, O, Na,  Mg, Al, Si, Ca, Ti, Ba, Eu/Fe] abundance ratios. Some default values are [C/Fe]=-0.2,  [N/Fe]=+1.0, and [O, Mg, Si, Ca, Ti, Eu/Fe] $=[\alpha$/Fe]. For the other elements, literature or fitted abundances are included, otherwise a solar abundance ratio is assumed.
 
 \item Spectral resolution, FWHM, the wavelength range being sampled, and the spectral resolution of the computed spectra $\delta \lambda$. The default values are FWHM~$= 0.2$~\AA\ and $\delta \lambda = 0.1$~{\rm \AA}. In order to compare to the sample observed spectra, we used FWHM~$= 1.0$~{\rm \AA} and $\delta \lambda = 0.5$~{\rm \AA}.
 
\end{itemize}

 The synthetic stellar spectra are generated using the {\sc PFANT} code described in Barbuy (1982), Cayrel et al (1991), Barbuy et al. (2003) and Coelho et al. (2005), and updated in Barbuy et al. (2018b).Given a stellar model atmosphere and lists of atomic and molecular lines, the code computes a synthetic spectrum assuming local thermodynamic equilibrium (LTE). 
The atomic line list is that of VALD3 (Ryabchikova et al. 2015).
For some specific lines we used updated oscillator strenghts,
in particular for the Mg I triplet at 5167-5183 (Pehlivan Rhodin et al. 2017).
The molecular line list calibrated through several stellar spectroscopic studies is described in 
Barbuy et al. (2018b). The MARCS LTE atmospheric models by Gustafsson et al. (2008) are employed. 

For each star along the isochrone we compute the synthetic spectrum,
taking into account the abundances derived from high-resolution spectroscopy
initially, and if needed, further fitting abundances the lines given in Table \ref{tablines}.

C. Computation of the integrated spectra:

The spectra of the SSP are created using the following integral

 $$f(\lambda) = \int_ {m_1}^{m_2} s(\lambda, m) \phi(m) dm$$

\noindent where $s(\lambda, m)$ is the spectrum of an individual star with mass $m$ as a function of the wavelength $\lambda$ and $\phi(m)$ is the IMF. 

The code reads a list of the stellar parameters T$_{\rm eff}$, $log$ g, mass, and luminosity log (L/L$_{\odot}$), and
the corresponding synthetic stellar spectra S($\lambda$, mass), computed
for each pair of (T$_{\rm eff}$, gravity log~g). The stellar population is
then divided into N mass bins, with a lower mass m$_{1}$=0.08 M$_{\odot}$, and
m$_{2}$  is the highest mass found in the isochrones being fitted. 

The package that computes SSP spectra with variable chemical abundances
 is available under request.

The abundances are varied all together, which is important, due in particular  to effects of some of them
as electron donors. Fe, Al, $\alpha$-elements such as Ca, are 
electron donors, and if enhanced, they contribute to increase the
continuum by H-. For this reason, it is important to compute them
enhanced at the same time. Therefore,
our method is different from those by Conroy et al. (2018) or 
Trager et al. (1998), where
each element is enhanced separately; it is also different from
the method by Colucci et al. (2017)
where the integrated spectra are analysed by measuring EWs of lines.

Our method is more similar to that by Sakari et al. (2013) and 
Larsen et al. (2017, and references therein), consisting of
full spectrum fitting, with the difference of us fitting
individual selected lines for element abundance derivation.

\section{Reproduction of integrated spectra of sample clusters}

We have computed integrated spectra for the six sample GCs, using the
prescriptions described in the previous section. The overall fit in the wavelength
range 4500-9000 {\rm \AA} is very satisfactory, as can be seen in Figures 15-19 in the Appendix,
where the full spectrum of NGC 6522 is fit.
Residuals for regions of 100 {\rm \AA} were measured, and the resulting
ratios between the synthetic and observed spectra are below 2\%.

We further inspected suitable lines of interesting elements,
 for which we are able to fit
element abundances. The fits to the selected lines,
reported in Table \ref{tablines},  proceeded as
in high-resolution spectroscopy, i.e., by fitting the local
continuum, and finding the best abundance.
 For the strong lines (Table \ref{mgca}),  the abundance
is not changed, because the bottom of these lines cannot be fitted
in LTE, as discussed in Sect. 4.2.

\begin{table*}
\begin{center}
  \caption{List of Lines Inspected in the Integrated Spectra, and Resulting Abundances}
\label{tablines}
\begin{tabular}{lcccccccccc}
\tableline
\hline
\hbox{species} & \hbox{$\lambda$({\rm \AA})} & \hbox{$\chi_{\rm ex}$ (eV)} & \hbox{log gf} &
\hbox{NGC 6171} & \hbox{NGC 6522} & \hbox{NGC 6624} & \hbox{NGC 6637} & \hbox{NGC 6723} & \hbox{NGC 6752} & \\
\hline
Na I          & 5682.633  & 2.10       & $-$0.71   & 0.00$\pm$0.15 & 0.00 & 0.20  & 0.00 & --- & 0.00 & \\
Na I          & 5688.194  & 2.11       & $-$1.40   & 0.35$\pm$0.15 & 0.35 & 0.35  & 0.35 & 0.35 & 0.35 &   \\
Na I          & 5688.205  & 2.11       & $-$0.45   & 0.35$\pm$0.15 & 0.35 & 0.35  & 0.35 & 0.35 & 0.35 &    \\
Na I          & 6154.230  & 2.10       & $-$1.56   & 0.35$\pm$0.15 & 0.20 & 0.35  & 0.35 & 0.35 & --- &   \\
Na I          & 6160.753  & 2.10       & $-$1.26   & 0.00$\pm$0.15 & 0.00 & 0.35  & 0.00 & 0.35 & 0.35 &   \\
Na I          & 8183.256  & 2.10       & $-$0.47   & 0.35$\pm$0.15 & 0.20 & 0.00  & 0.00 & 0.20 & 0.20 &    \\
Na I          & 8194.790  & 2.10       & +0.24     & 0.20$\pm$0.15 & 0.20 & 0.20  & 0.00 & 0.00 & 0.35 &    \\ 
Mg I          & 5528.405  & 4.34       & $-$0.547  & 0.51$\pm$0.15 & 0.23 & 0.42 & 0.28 & 0.23 & 0.38 &      \\
Mg I          & 5711.088  & 4.34       & $-$1.842  & 0.51$\pm$0.10 & 0.23 & 0.42 & 0.28 & 0.45 & 0.25 &      \\
Al I          & 6696.185  & 4.02   & $-$1.58   & 0.25$\pm$0.20 & 0.20 & 0.30 & 0.20 & 0.31 & 0.10 &    \\
Al I          & 6696.204  & 4.02   & $-$1.58   & 0.25$\pm$0.20 & 0.20 & 0.30 & 0.20 & 0.31 & 0.10 &    \\
Al I          & 6696.788  & 4.02       & $-$1.42   & 0.25$\pm$0.20 & 0.20 & 0.30 & 0.20 & 0.31 & 0.10 &    \\
Al I          & 6698.673  & 3.14       & $-$1.65   & 0.25$\pm$0.20 & 0.20 & 0.30 & 0.20 & 0.31 & 0.10 &    \\ 
Si I          & 6414.99   & 5.87  & -1.13  & 0.47$\pm$0.20 & -0.05 & 0.15 & 0.45 & -0.18 & 0.30 &  \\
Si I          & 7405.79   & 5.61  & -0.66  & 0.05$\pm$0.20 & -0.50 & 0.08 & 0.05 & 0.20  & 0.47 &  \\
Si I          & 7415.96   & 5.61  & -0.73  & 0.15$\pm$0.20 & 0.20  & 0.38 & 0.25 & 0.20  & 0.47 &  \\
Si I          & 7423.51   & 5.62  & -0.58  & 0.05$\pm$0.20 & -0.18 & 0.08 & 0.05 & -0.05 & 0.0  &  \\
Ca I          & 6161.295  & 2.51       & $-$1.02   & 0.21$\pm$0.15 & 0.13 & 0.45  & 0.40 & 0.53 & 0.20 & \\
Ca I          & 6162.167  & 1.89       & $-$0.09   & 0.21$\pm$0.15 & 0.13 & 0.45  & 0.40 & 0.53 & 0.20 &            \\
Ca I          & 6439.080  & 2.52       & +0.3      & 0.21$\pm$0.15 & 0.13 & 0.45  & 0.40 & 0.53 & 0.20 &      \\
Ti I         & 5965.825  & 1.88      & $-$0.42   & -0.04$\pm$0.10 & 0.04 & 0.37 & 0.24 & 0.24 & 0.20 &         \\ 
Ti I          & 6261.106  & 1.43      & $-$0.48   & -0.04$\pm$0.10 & 0.10 & 0.37 & 0.24 & 0.50 & 0.20 &           \\
Ti I          & 6336.113  & 1.44      & $-$1.74   & -0.04$\pm$0.10 & 0.04 & 0.37 & 0.24 & 0.24 & 0.20  &           \\
Ba II         & 6496.90   & 0.604      & -0.32        & 0.00$\pm$0.15 & 0.02 & 0.10 & 0.22 & 0.22  & 0.00 &          \\
Eu II         & 6645.064  & 1.38       & +0.12       & 0.50$\pm$0.15 & 0.55 & 0.00 & 0.45 & 0.00 & 0.50 &        \\
\tableline
\multicolumn{11}{l}{Note. Uncertainties given for NGC 6171 are the same for all clusters for each respective line.}
\end{tabular}
\end{center}
\end{table*}

\subsection{Lines of Mg I, Na I, Al I, Si I, Ca I, Ti I, Ba II, Eu II }

In this subsection we present the resulting abundances from
spectrum fitting for best lines for the elements 
Mg I, Na I, Al I, Si I, Ca I, Ti I, Ba II, and Eu II. The 
results are adopted as the element abundance in the
sample clusters. The reason we prefer weaker lines
rather than strong lines to derive abundances is the simple
fact that in that case the abundance increases linearly
with EW (e.g. Gray 2005).

The fits to lines for all six clusters, for chosen lines
of Na, Mg, Al, Si, Ca, Ti, Ba, and Eu, are described below.
The element abundances obtained or adopted
are given in the panels and gathered
in Table \ref{tablines}. The Na lines and abundances are
treated separately as reported in Section 5.

The uncertainties in the abundances derived in Table \ref{tablines}
are related to the strength of the line, and surrounding
blends allow or not a good continuum fit. The uncertainties
given in the Table for NGC 6171 are adopted for all clusters
for the same respective lines.
 
\paragraph{Mg Lines}
 
\begin{figure*}
\centering
\includegraphics[width=0.75\textwidth]{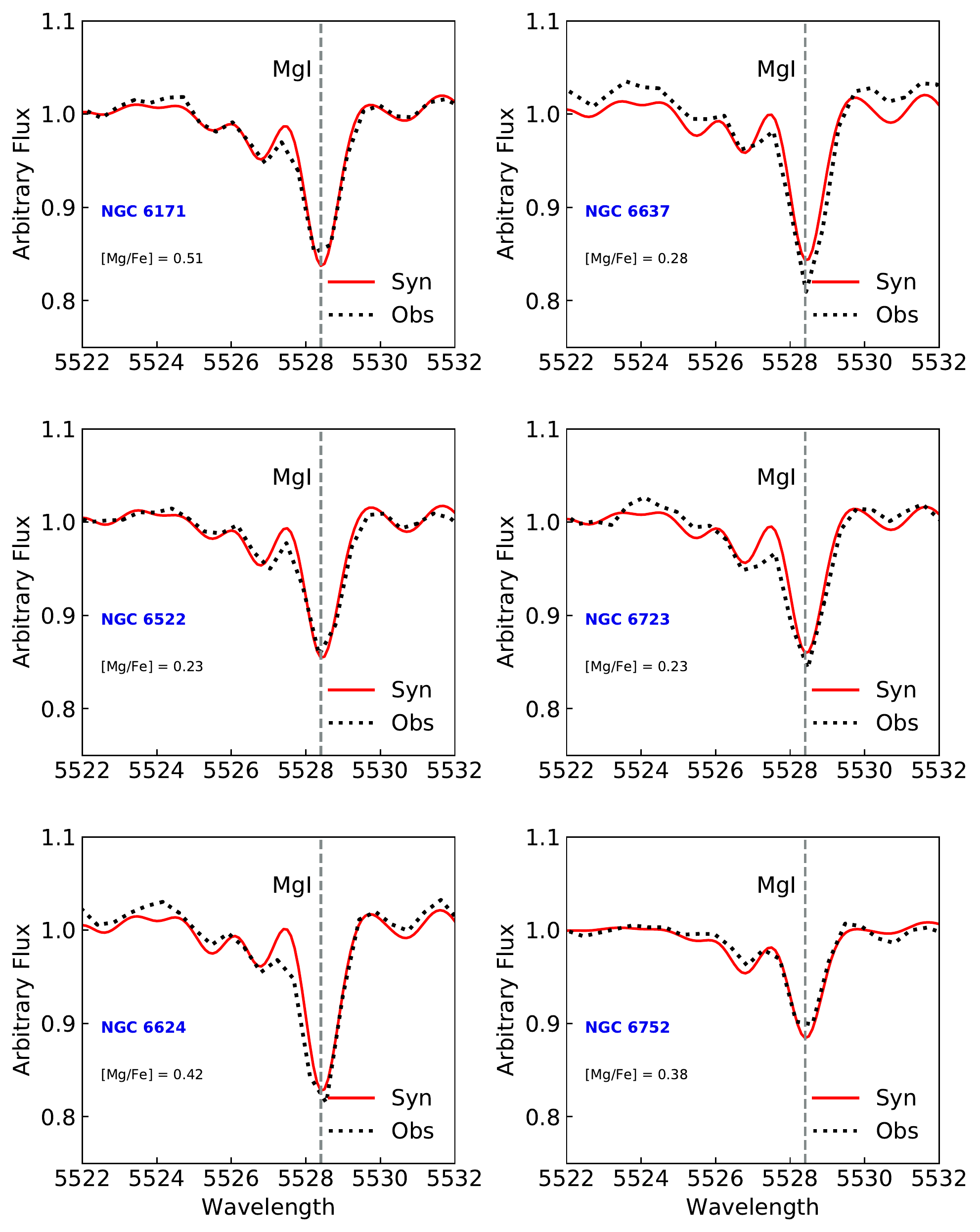}
\caption{The observed and synthetic spectra for the Mg I line in 5528.405 {\rm \AA}.}
\label{fig:All_MgI_5528}
\end{figure*}

\begin{figure*}
\centering
\includegraphics[width=0.75\textwidth]{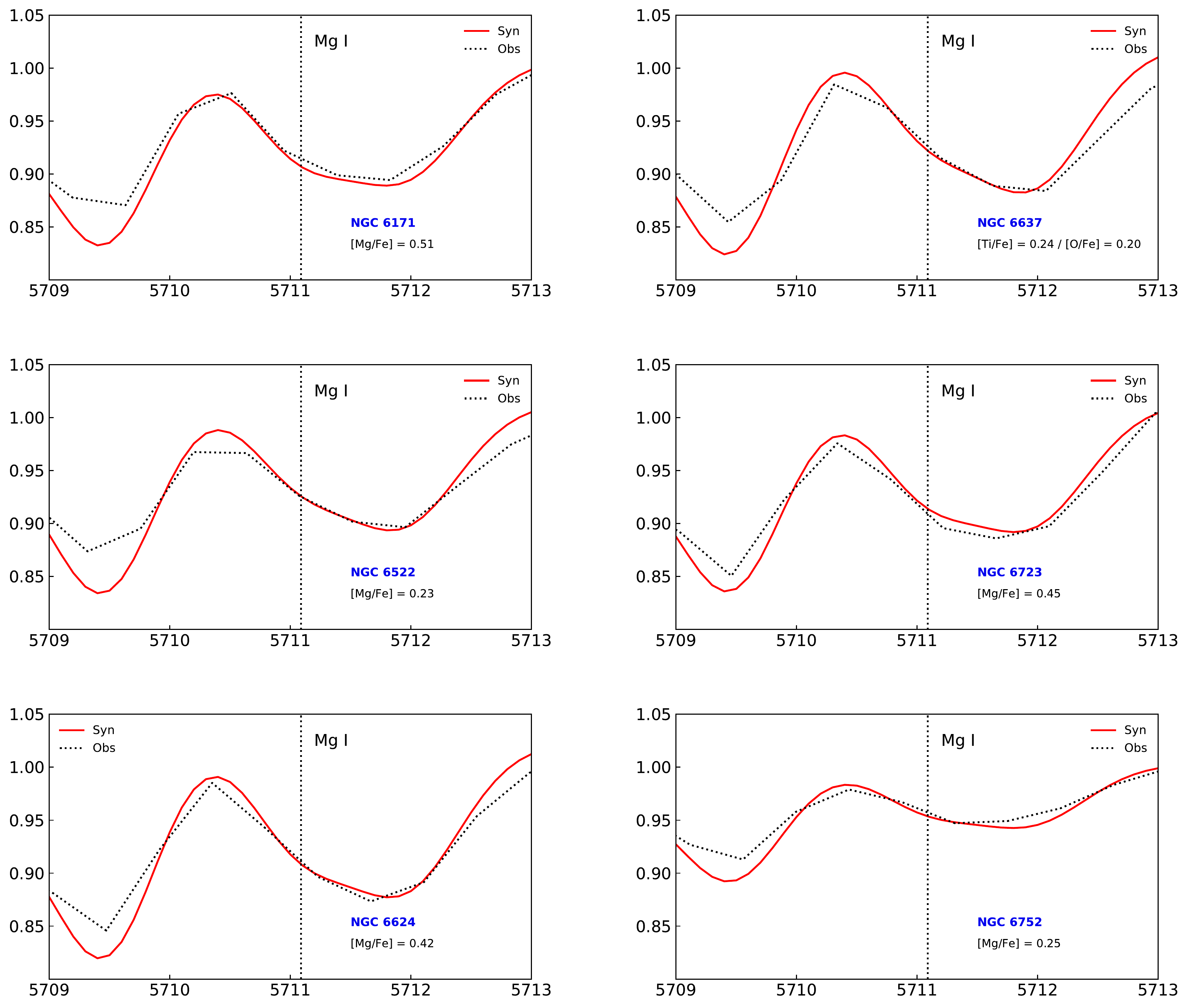}
\caption{The Mg 5711.088 {\rm \AA} line for the sample clusters.}
\label{fig:All_MgI_5711}
\end{figure*}

Figures \ref{fig:All_MgI_5528} and \ref{fig:All_MgI_5711} show the
Mg I 5528.405 {\rm \AA} and  5711.08 {\rm \AA} lines that are well-fitted for the six clusters.

\paragraph{Al Lines}

\begin{figure*}
    \centering
    \includegraphics[width = 0.68\textwidth]{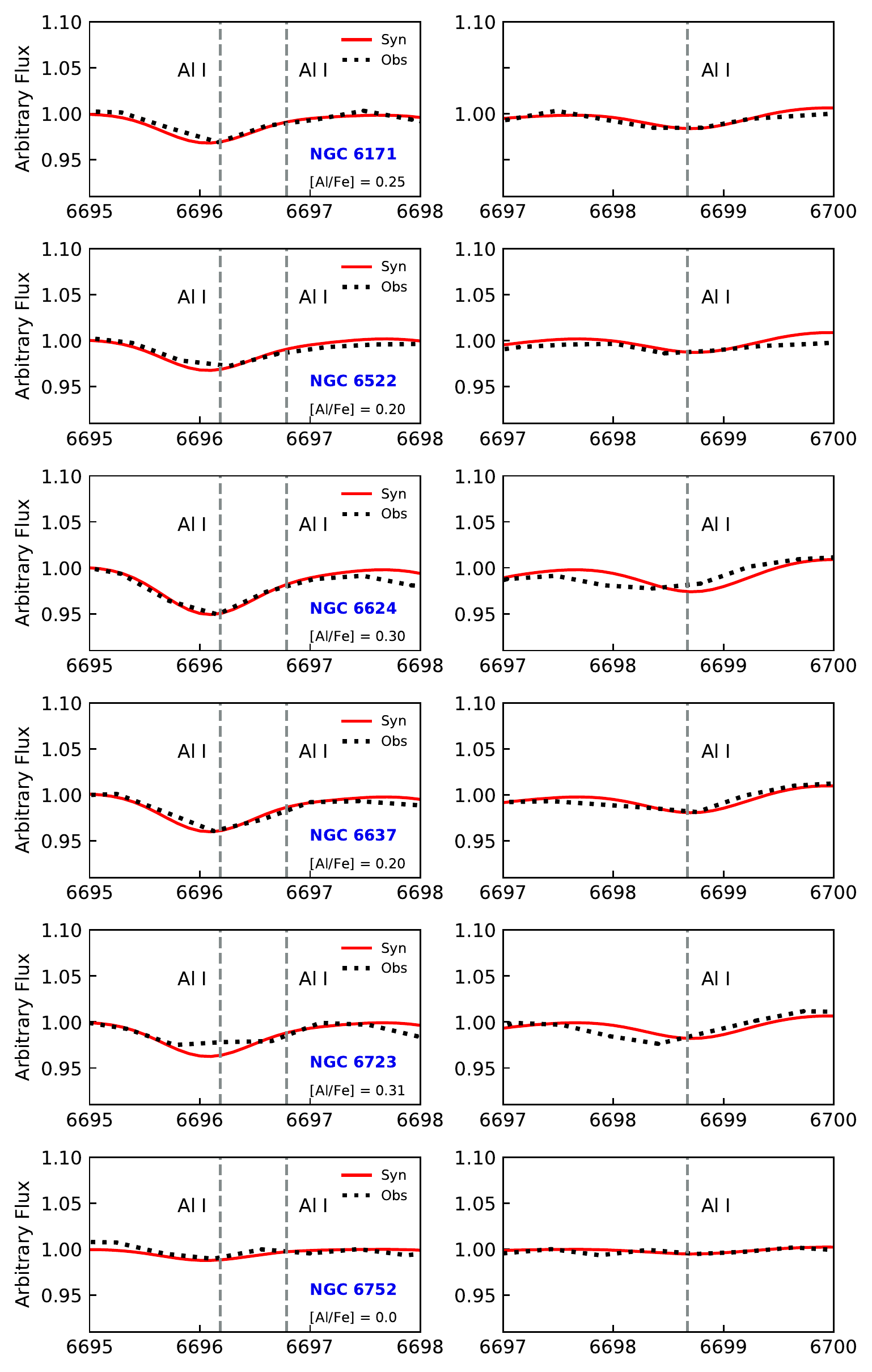}
    \caption{Observed and synthetic spectra in region of the Al I lines.}
    \label{fig:Al}
\end{figure*}

The Al I 6696.185, 6696.788, 6698.673 {\rm \AA} lines are
shown in Figure \ref{fig:Al}. The calculations are in a good agreement 
with the integrated spectra, where  for the clusters NGC 6723,
NGC 6752 and NGC 6522, solar ratios are adopted.
For NGC 6171 there are no Al abundances available in the literature. 
 NGC 6171, NGC 6624 and NGC 6637 show Al enhancements,
 that are also found in bulge field stars, where it behaves as an $\alpha$-element
 (see McWilliam 2016, Barbuy et al. 2018a).
 Note that Al enhancements can also be an indicator of a second generation of stellar populations
 in globular clusters (e.g. Renzini et al. 2015, Bastian \& Lardo 2018).

\paragraph{Si Lines}

\begin{figure*}
 \centering
 \includegraphics[width = 0.73\textwidth]{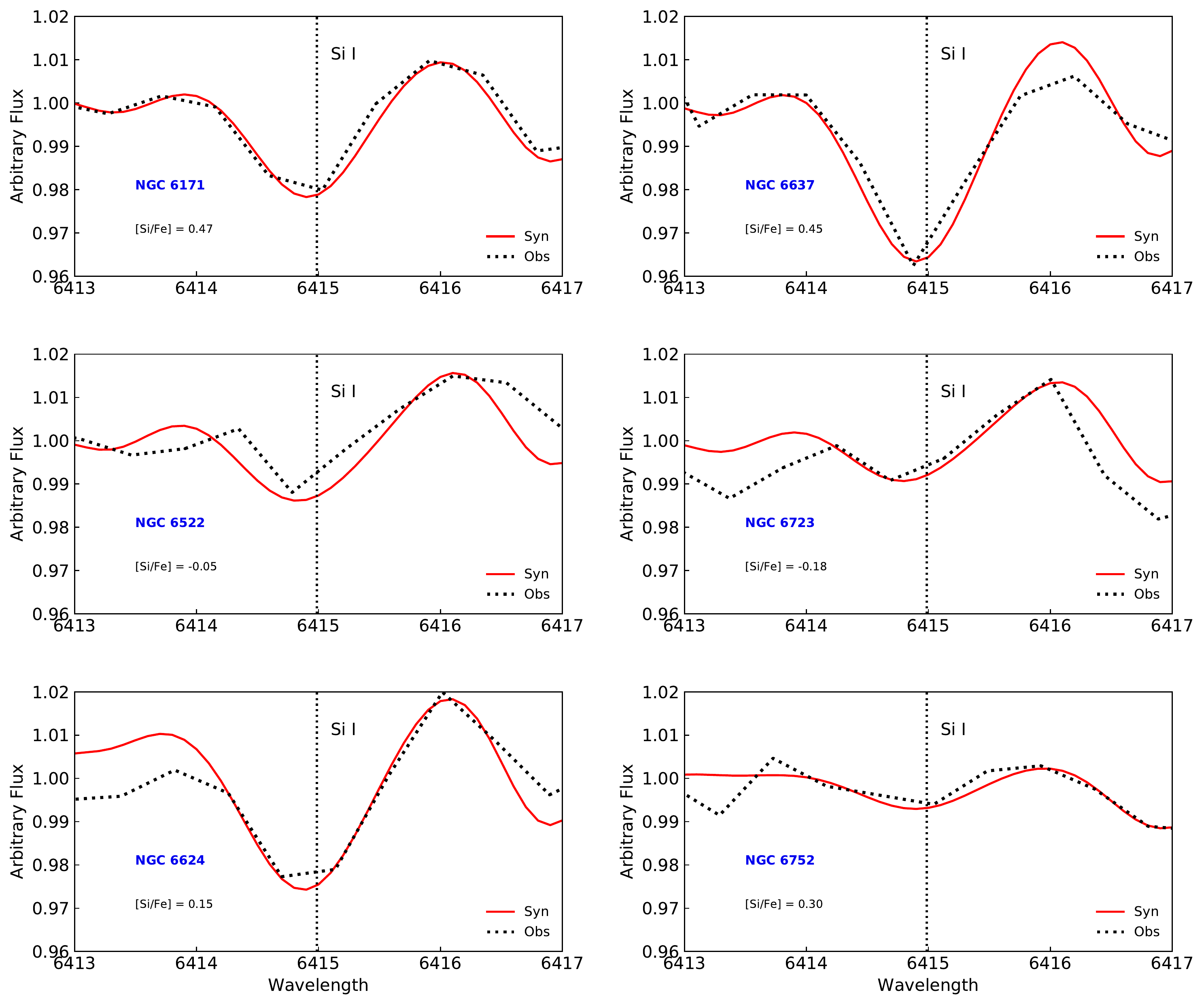}
 \caption{The synthetic and observed spectra for the six clusters of our sample in the region of the 6414.987 {\rm \AA} Si line}
 \label{fig:Si_6414_All}
\end{figure*}

\begin{figure*}
 \centering
 \includegraphics[width = 0.9\textwidth]{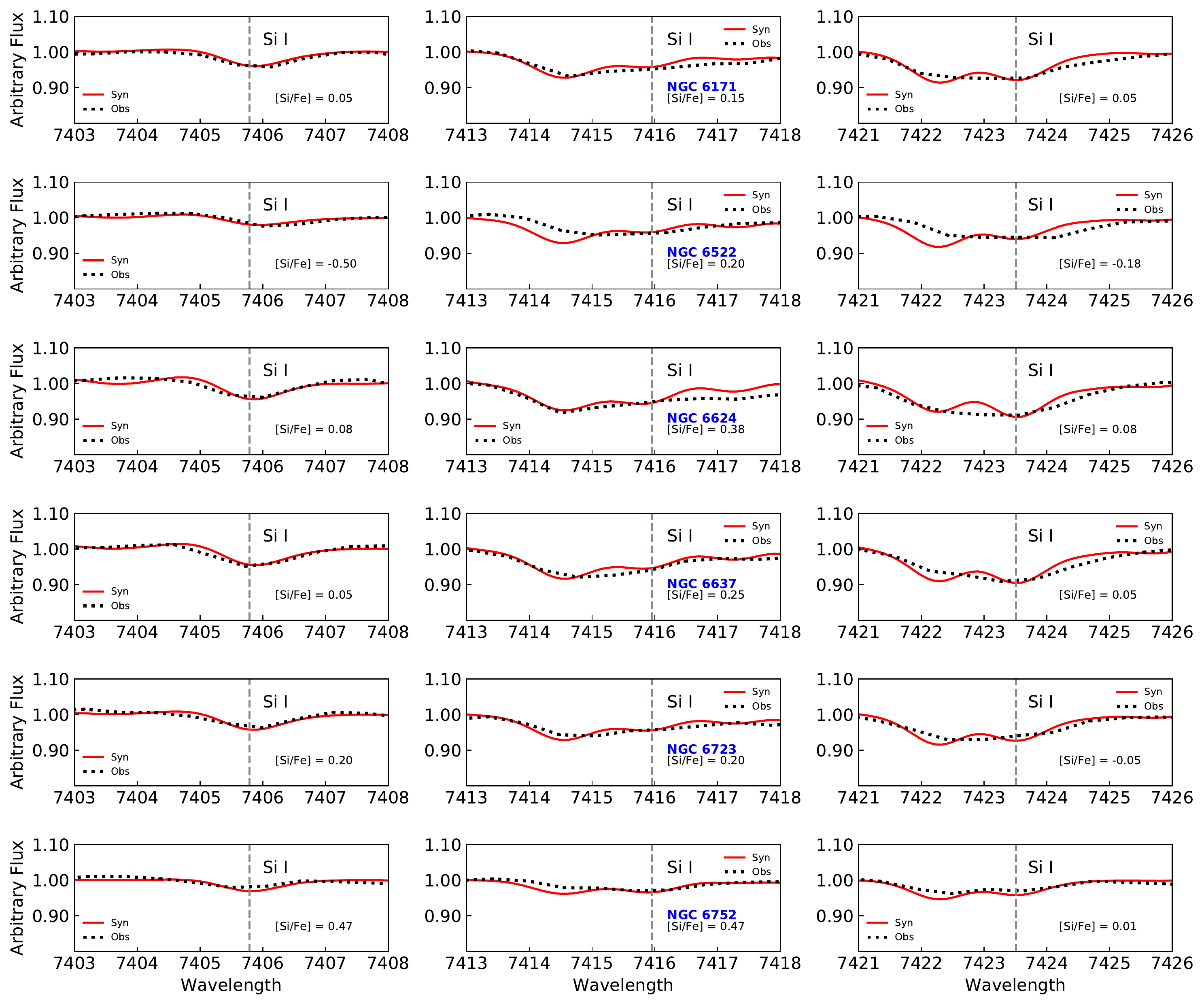}
 \caption{The 7405.79, 7415.9, and 7423.51 {\rm \AA} Si Lines in the synthetic and observed spectra for our sample.}
 \label{fig:Si_7405_All}
\end{figure*}

Figures \ref{fig:Si_6414_All} and \ref{fig:Si_7405_All} show the fits to the
Si I 6414.99 and 7405.79 {\rm \AA} lines.

\paragraph{Ca Lines}

\begin{figure*}
    \centering
    \includegraphics[width = 0.75\textwidth]{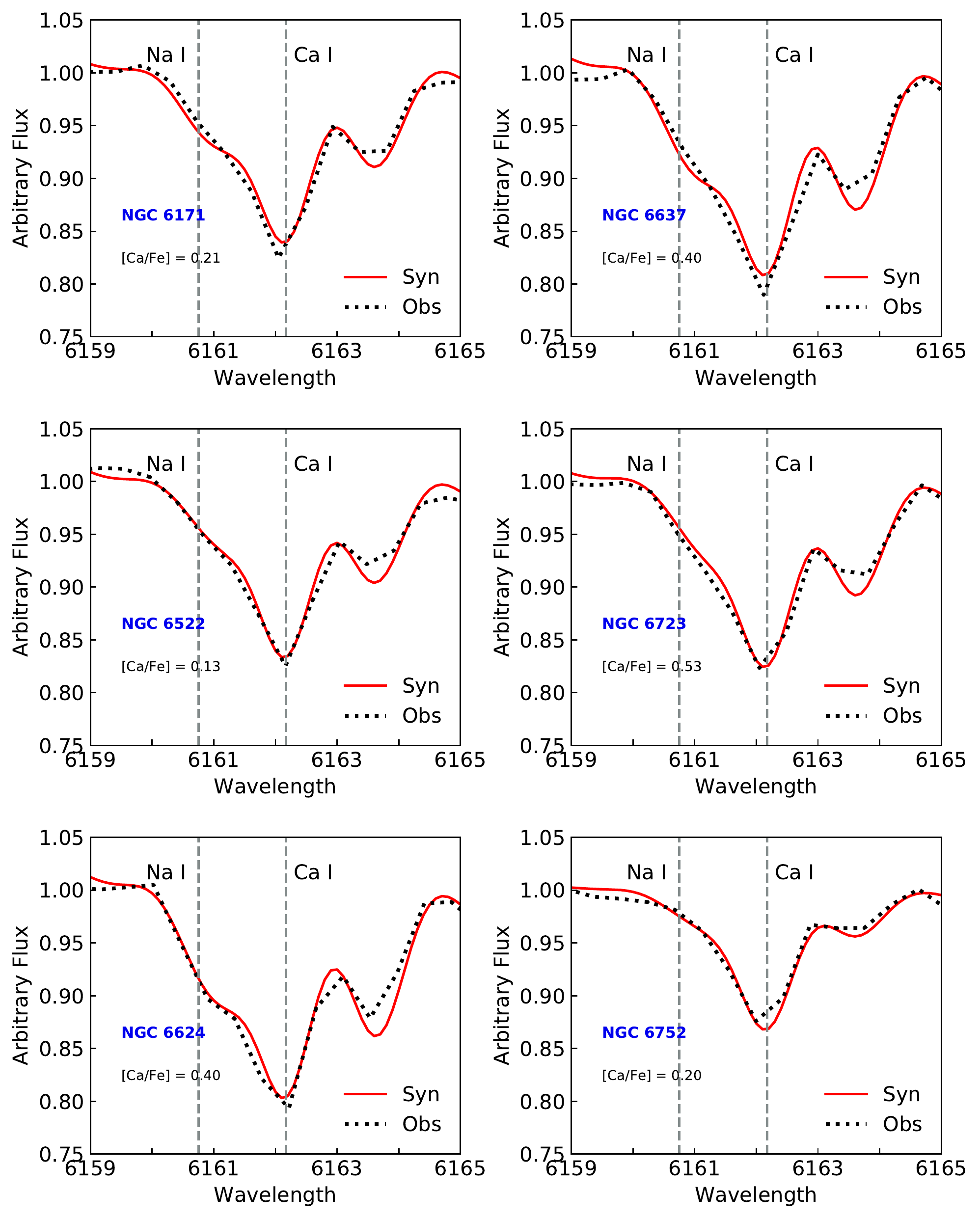}
    \caption{Observed and synthetic spectra in
      the region of the Ca I 6162.7 {\rm \AA} line.}
    \label{fig:All_Ca1_6162}
\end{figure*}

Figure \ref{fig:All_Ca1_6162} shows the fits to the Ca I 6162.167 {\rm \AA}, as well as to
the nearby Na I 6160.753 {\rm \AA} line, with the adopted abundances indicated
in the figure and in Table \ref{tablines}.
Usher et al. (2019) derived the Ca abundances for their full sample clusters, 
based on measurements of equivalent widths with well-defined continua and 
wavelength range of Ca II triplet lines (CaT). The values from Usher et al. (2019) and
the present results are compatible (See Section 4.3).
It is important to note that in this subsection we derived the Ca
abundance from weak to medium
lines, which are more suitable for abundance derivation, therefore a method
different from Usher et al. (2019), even if using the same spectra.
It is reassuring to conclude that the Ca abundances from the CaT strong lines
measured as indices, and the fit to normally weak to medium lines in the
integrated spectra, give very close results.

\paragraph{Ti Lines}

\begin{figure*}
 \begin{tabular}{ll}
  \includegraphics[width=0.45\textwidth]{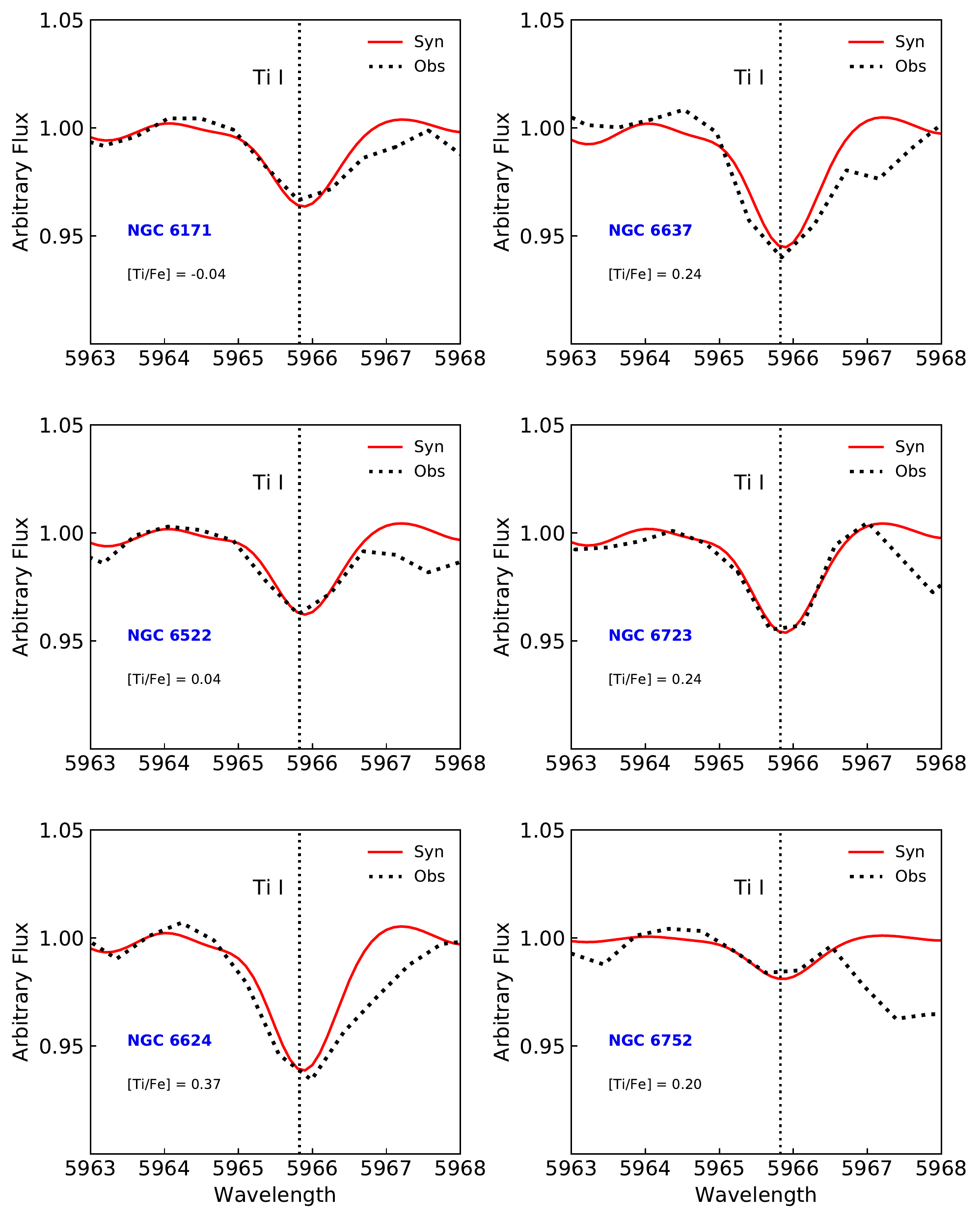} &
  \includegraphics[width=0.45\textwidth]{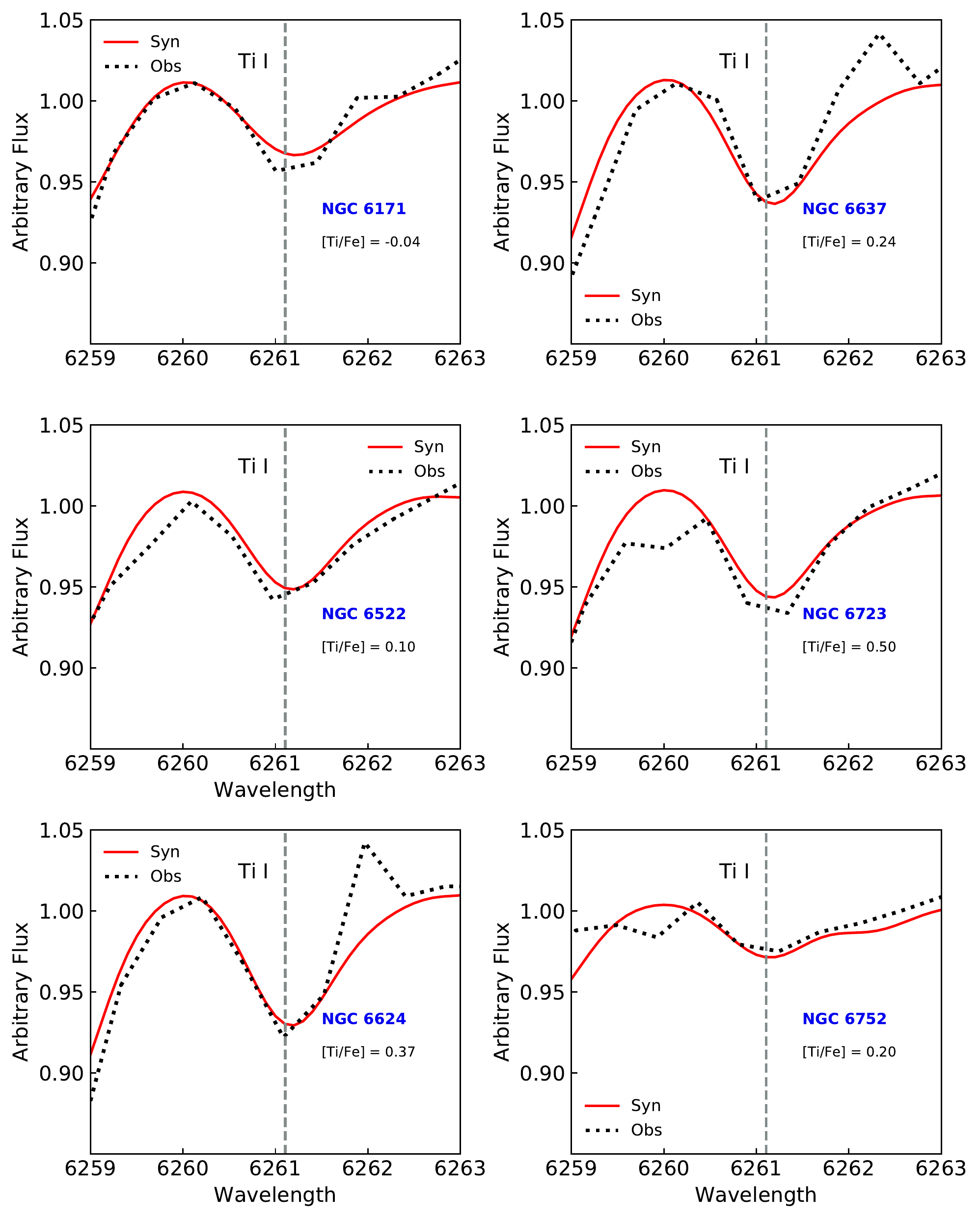}
 \end{tabular}
\caption{Fitting of synthetic to observed spectra for different regions with Ti I lines.}
\label{figAll_TiI}
\end{figure*}

Ti I 5965.825, 6261.106, and 6336.113 {\rm \AA} are used to derive Ti abundances,
with the first two lines shown in Fig. \ref{figAll_TiI}.

\begin{figure*}
    \centering
    \includegraphics[width = 0.70\textwidth]{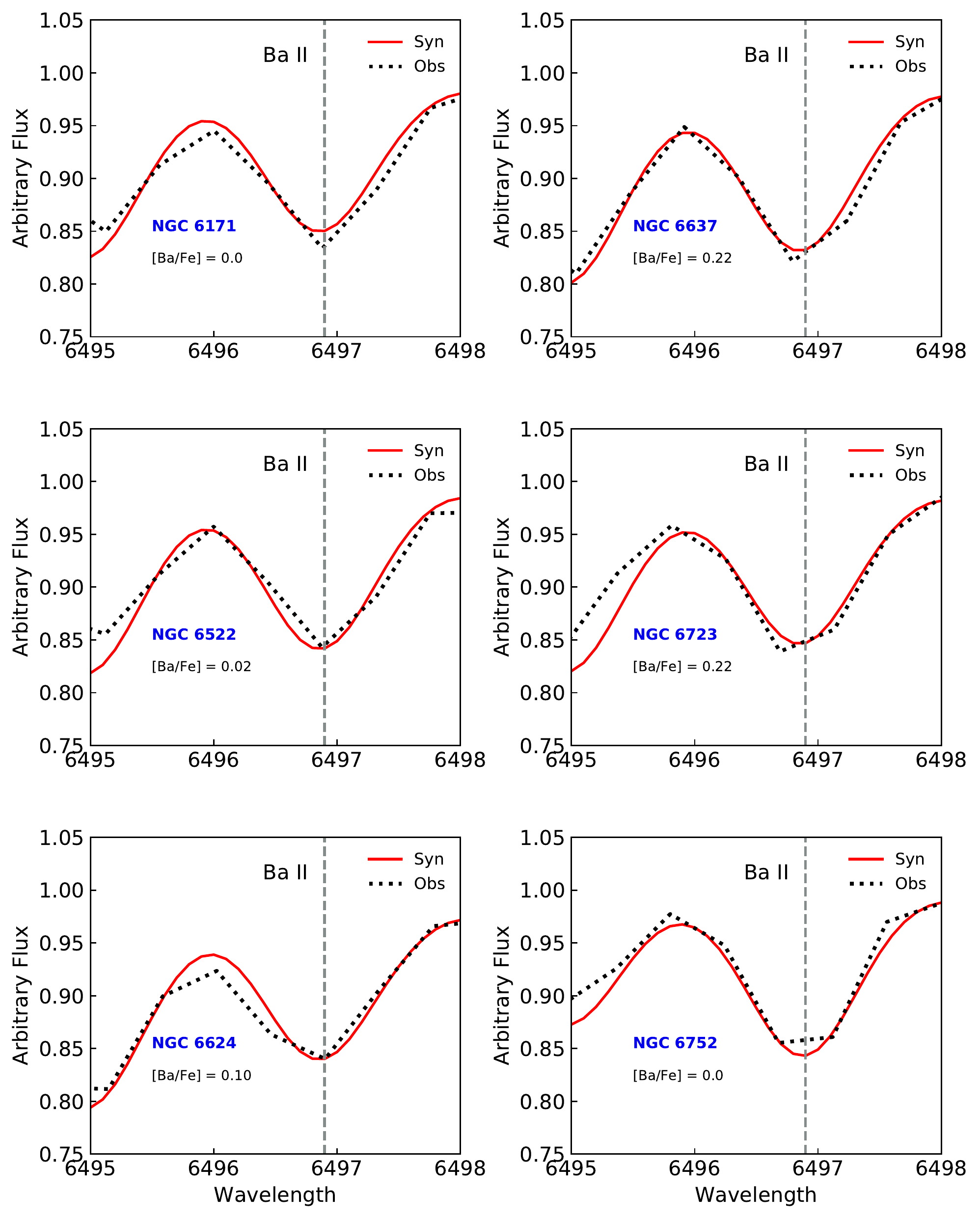}
    \caption{Observed and synthetic spectra in region of the line 6496 {\rm \AA} Ba II.}
    \label{fig:All_Ba2_6496}
\end{figure*}

\paragraph{Ba lines}
Fig. \ref{fig:All_Ba2_6496} shows the fits to the Ba II 6496.897 {\rm \AA} line,
with abundances indicated in the panels and in Table \ref{tablines}.
Hyperfine structure (HFS) is taken into account following Barbuy et al. (2014).

\begin{figure*}
    \centering
    \includegraphics[width = 0.7\textwidth]{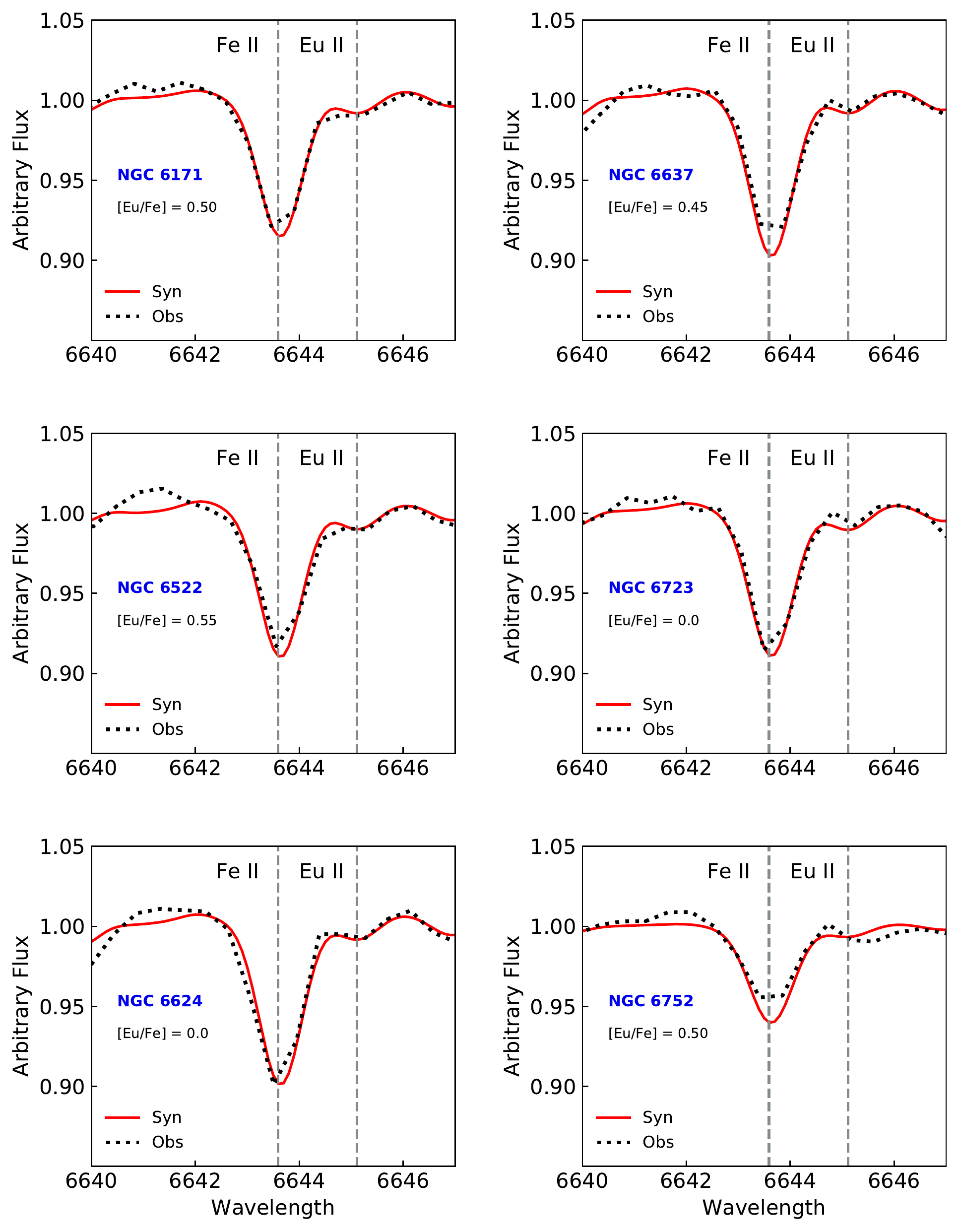}
    \caption{The 6645 {\rm \AA} Eu II line in the synthetic and observed spectra for each globular cluster of our sample.}
    \label{fig:All_Eu2}
\end{figure*}

\paragraph{Eu Line}
Figure \ref{fig:All_Eu2} gives the Eu II 6645.064 {\rm \AA} line well-fitted with
literature abundances when available (NGC 6522 and NGC 6637), and for the others
the line was fitted, with the resulting Eu abundance indicated in the figures,
and in Table \ref{tablines}. HFS is taken into account.

\subsection{Strong Mg I and Ca II triplet lines}\label{sec:Triplet}

For the strong lines of the Mg I triplet at 5167-5183 {\rm \AA},
and the Ca II triplet at 8498-8662 {\rm \AA}, we have adopted
oscillator strengths from the recent literature, and
derived astrophysical values for the damping constants to
reproduce their wings, applied to the Sun and Arcturus.
The abundances are adopted from the lines reported in Table \ref{tablines}.

The main mechanism of collisional broadening for metallic
lines in stars of spectral types F and M is due to
collisions with neutral hydrogen atoms.
The damping constant $\gamma$ for each atmospheric layer relates
to the interaction constant C$_6$ as follows (e.g. Gray 2005):

$$\gamma_{6}/N_{H} = 17v^{3/5}C_6^{2/5}$$

\noindent where v is the relative velocity between the colliding particles 
and N$_{\rm H}$ is the density of hydrogen atoms.

The radiative broadening (taken as standard for most lines with
$\gamma_{\rm R} = 2.21E15/\lambda^{2}$), as well as electron broadening are
also taken into account, to form the final Voigt profile
(Gray 2005). The values of $\gamma_{\rm R}$ and  $\gamma_{\rm e}$
are adopted from the VALD3 (Ryabchikova et al. 2015) and
Kur\'ucz webpage\footnote{http://www.cfa.harvard.edu/amp/ampdata/kurucz23/sekur.html},
and these are further described for example in Smith \& Drake (1988) and
Chmielewski (2000).

The adopted values of the interaction constant C$_6$ were obtained
by adopting the recent literature oscillator strengths, and fitting
of the wings. This is done by using the MARCS models
(Gustafsson et al. 2008) applied
to the Sun (Kur\'ucz 2005)\footnote{kurucz.harvard.edu/sun.html} and Arcturus (Hinkle et al. 2000).
We adopted the stellar parameters
 effective temperature (T$_{\rm eff}$), surface gravity (log~g), 
 metallicity ([Fe/H]) and microturbulent velocity (v$_{\rm t}$)  of
 (5770 K, 4.44, 0.0, 1.0 km.s$^{-1}$) for the Sun, and
 (4275 K, 1.55, -0.54, 1.65 km.s$^{-1}$) for Arcturus from 
Mel\'endez et al. (2003). We adopted 
abundances of A(Fe)=7.50 for the Sun, and
[Mg/Fe]=+0.15 and [Ca/Fe]=+0.10 for Arcturus. 
Note that the damping constant through the interaction constant
C$_6$ that fit the wings of strong lines, depends on the atmospheric
model grid adopted, as pointed out in Barbuy et al. (2003, their Fig. 1).

Another comment as regarding strong lines is that LTE
calculations do not reproduce the bottom of these lines,
given that they form in NLTE, and besides their formation
takes place in chromospheric layers, that are not included
in the usual model atmospheres.

\begin{figure*}
\centering
\includegraphics[width=0.75\textwidth]{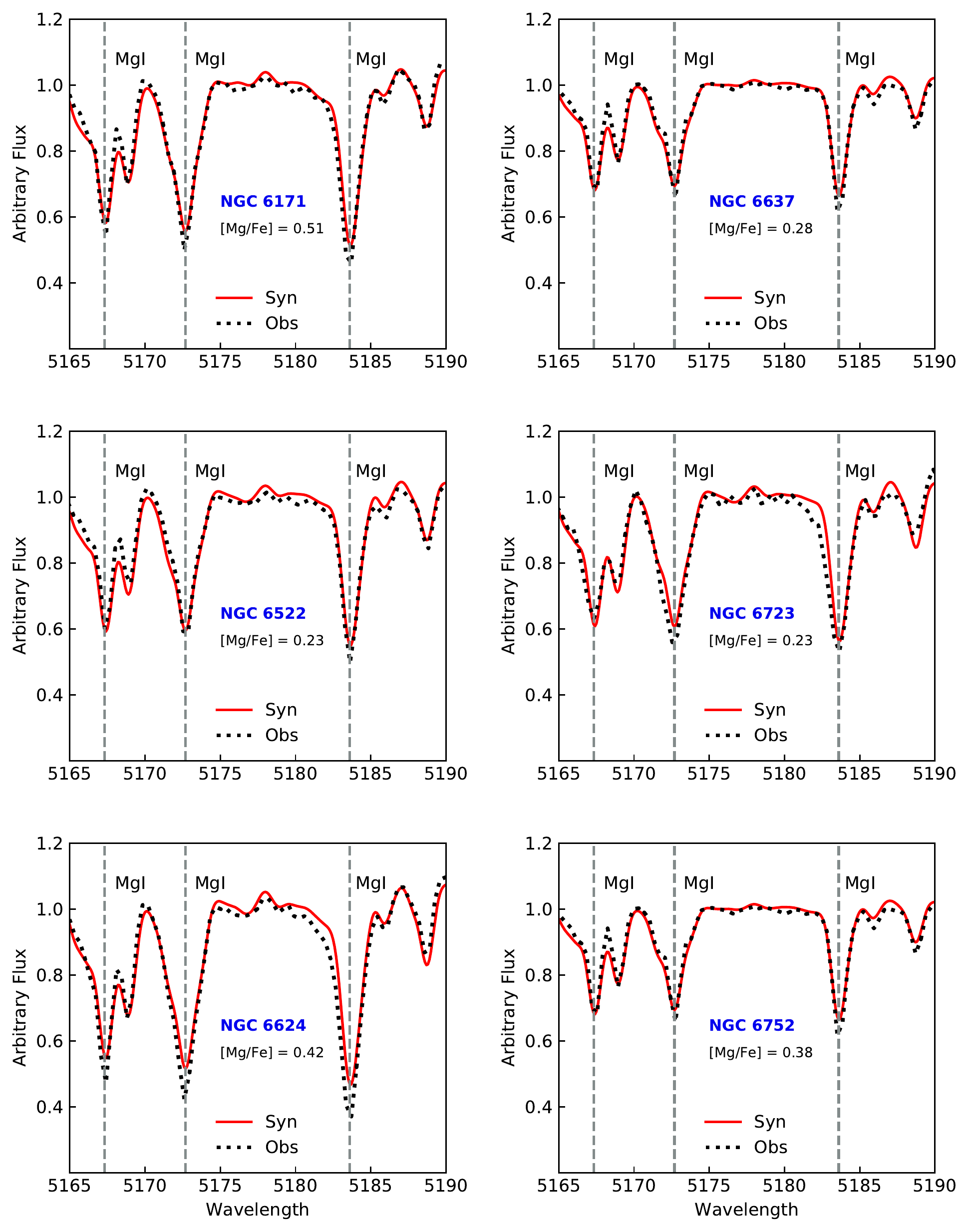}
\caption{The Mg I triplet lines at 5167.3, 5172.68 and 5183.60 {\rm \AA} for the sample clusters.}
\label{fig:All_MgI_triplet}
\end{figure*}

\begin{figure*}
    \centering
    \includegraphics[width = 0.95\textwidth]{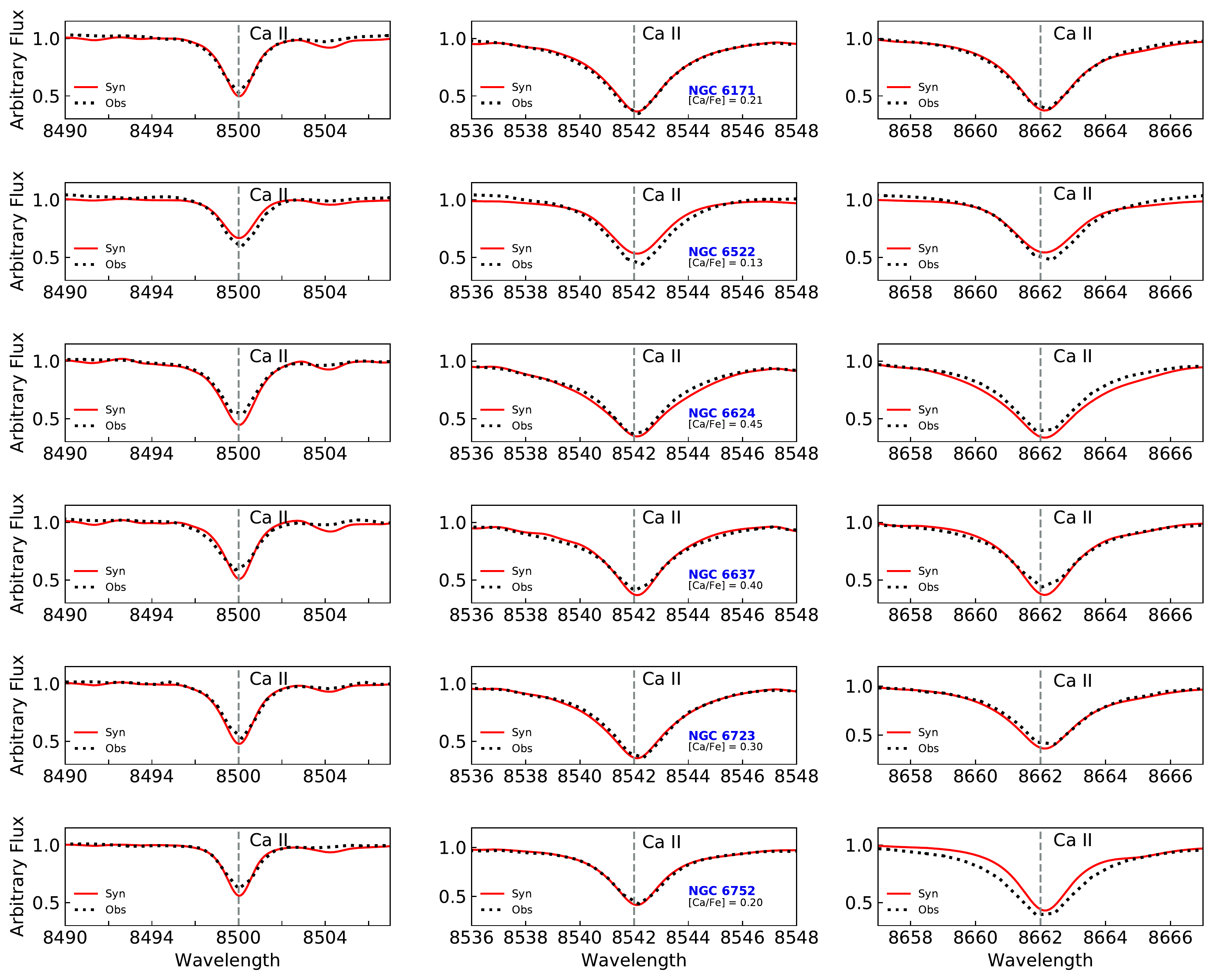}
    \caption{Ca II triplet lines for the sample clusters.}
    \label{fig:CaT}
\end{figure*}

For the Mg I triplet lines at 5167.3, 5172.68 and 5183.60 {\rm \AA}
and Ca II triplet lines (CaT) at 8498.023, 8542.091, and 8662.141 {\rm \AA},
we used the atomic constants listed in Table \ref{mgca}. This resulted in very good fits
 shown in Figures \ref{fig:All_MgI_triplet} and \ref{fig:CaT}.

\begin{table*}
\begin{center}
\caption{Atomic Constants Adopted for the Mg I and Ca II Triplet Lines}
\label{mgca}
\setlength{\tabcolsep}{10pt} 
\begin{tabular}{lccccccc}
\tableline 
\hline
\hbox{species} & \hbox{$\lambda$({\rm \AA})} & \hbox{$\chi_{\rm ex}$ (eV)} & \hbox{log gf} &
C$_6$ & $\gamma_{R}$/NH & $\gamma_{e}$/NH & \\
\hline
MgI & 5167.3216 & 2.7091 &$-$0.854$^1$ & 0.3E-29   & 1.0E+08     & 2.0E-07&  \\ 
MgI & 5172.684 & 2.7116  & $-$0.363$^1$ &  idem   & idem      & idem &        \\
MgI & 5183.604 & 2.7166  & $-$0.168$^1$ &  idem    & idem      & idem &      \\
CaII & 8498.023 & 1.6924 & $-$1.312$^2$    & 0.9E-32    & 3.6E+08     & 2.6E-07 & \\
CaII & 8542.091 & 1.6999 & $-$0.362$^2$    & 0.8E-32  & 3.0E+08     & idem &   \\
CaII & 8662.141 & 1.6924 & $-$0.623$^2$    & 0.8E-32   & 2.95E+08     & idem &  \\
\tableline
\multicolumn{7}{l}{Note. References: 1 Phelivan Rhodin et al. 2017, 2 Kur\'ucz (2005).}
\end{tabular}
\end{center}
\end{table*}

\subsection{Literature abundances}

The resulting abundances derived in the above subsections
are then compared with other methods
of abundance derivation from integrated light reported in the
literature. 
The element abundances derived from different methods and sources
are compiled in Table \ref{tab2}. The method employed is identified as
high-resolution spectroscopy (HRS), CMD,
compilation (comp.),
and from integrated spectra: equivalent widths (ISEW), 
full spectrum fitting (ISFF), or finally the fit of individual lines
(ISfit), the latter being our adopted method.

Below we briefly describe methods employed in the literature
for derivation of abundances applied to the
six sample GCs. The references
and comments given here are by no means exhaustive, since
there is more extensive literature addressing the clusters.

Usher et al. (2019) derived Ca abundances from EW measurements
following prescriptions given in Usher et al. (2012), similarly
to other studies in the literature such as
e.g. V\'asquez et al. (2018) applied to individual stars,
or Foster et al. (2010) applied to integrated light.
Usher et al. (2019) derive Ca abundances for all sample clusters in the
present work and their values are given in Table \ref{finalabunds}.

Colucci et al. (2017), following earlier work by Sakari et al. (2013),
measured EWs of individual lines 
of Mg, Ca, Ti, Cr, Ba, and Eu, from
the same WAGGS spectra used here, and they report 
the corresponding element abundances. 
They also use spectrum synthesis to fit iron peak and heavy
element lines
that require inclusion of HFS.
The cluster NGC 6752 is a common factor in Colucci et al. (2017) and the present work, noting that
in most cases Colucci et al. chose different lines relative to
our Table \ref{tablines}. In Table \ref{colucci} we compare
the abundances for lines in common between Colucci et al. and
the present work, for NGC 6752.

Larsen et al. (2017) obtained high-resolution integrated spectra
(R$\sim$40,000) of seven clusters, including NGC 6752 in
common with the present work. Their method of full 
spectrum synthesis, as well as the use of the
Dartmouth isochrones is similar to that of the present work.
They apply overall fitting to all the spectrum, finding
the best abundances to fit it all, therefore in this respect,
it is different from the present work where we have chosen
the most reliable lines to be fitted.

Conroy et al. (2018) present a full spectrum fitting method
to derive abundances from integrated spectra. They use
Dartmouth isochrones and MILES empirical spectra 
(S\'anchez-Bl\'azquez et al. 2006), complemented by
a library of spectra in the near-infrared. An important
addition is the calculation of synthetic spectra 
with variable abundances, in order to adapt the observed
spectra to response functions due to variable abundances.
From fitting spectra by Schiavon et al. (2005), they
derive abundances of Mg, Si, Ca, and Ti for 41 globular
clusters, including our six sample clusters.

\begin{table*}
\begin{center}
\caption{Comparison of abundances Line-by-line with Colucci et al. (2017)
for NGC 6752, assuming their [Fe/H] = -1.58}
\label{colucci}
\setlength{\tabcolsep}{10pt} 
\begin{tabular}{lcccccccc}
\tableline
\hline
\hbox{species} & \hbox{$\lambda$({\rm \AA})} &  \hbox{A(X)$_{\rm \odot Colucci}$} &
 \hbox{A(X)$_{\rm Colucci}$} & \hbox{[X/Fe]$_{\rm Colucci}$} & \hbox{[X/Fe]$_{\rm this work}$} &\\
\hline
NaI          & 5682.633  & 6.44 & 4.95 & +0.09 & +0.00  & \\
NaI          & 5688.200  & 6.34 & 4.75 & -0.01 & +0.35  &  \\ 
MgI          & 5528.405  & 7.52 & 6.05 & +0.11   & +0.38  &     \\
MgI          & 5711.088  & 7.58 & 5.95 & -0.05  &  +0.25 &   \\
SiI          & 7405.79   & 7.56 & 6.56 & +0.58 &  +0.47 &   \\
SiI          & 7415.96   & 7.63 & 6.46 & +0.41 &  +0.47 &   \\
SiI          & 7423.51   & 7.60 & 6.26 & +0.24 &  +0.00 &   \\
CaI          & 6162.167  & 6.40 & 5.06 & +0.24 & +0.20 &         \\
CaI          & 6166.440  & 6.36 & 5.16 & +0.38 & +0.20 &          \\
CaI          & 6439.080  & 6.02 & 4.96 & +0.52 & +0.20 &   \\
TiI          & 5866.449  & 5.04 & 4.00 & +0.54 & +0.20 &       \\ 
BaII         & 6496.90   & 2.18 & 0.73 & +0.13  & 0.00  &        \\

\tableline
\end{tabular}
\end{center}
\end{table*}

\begin{figure*}
    \centering
    \includegraphics[width = 0.75\textwidth]{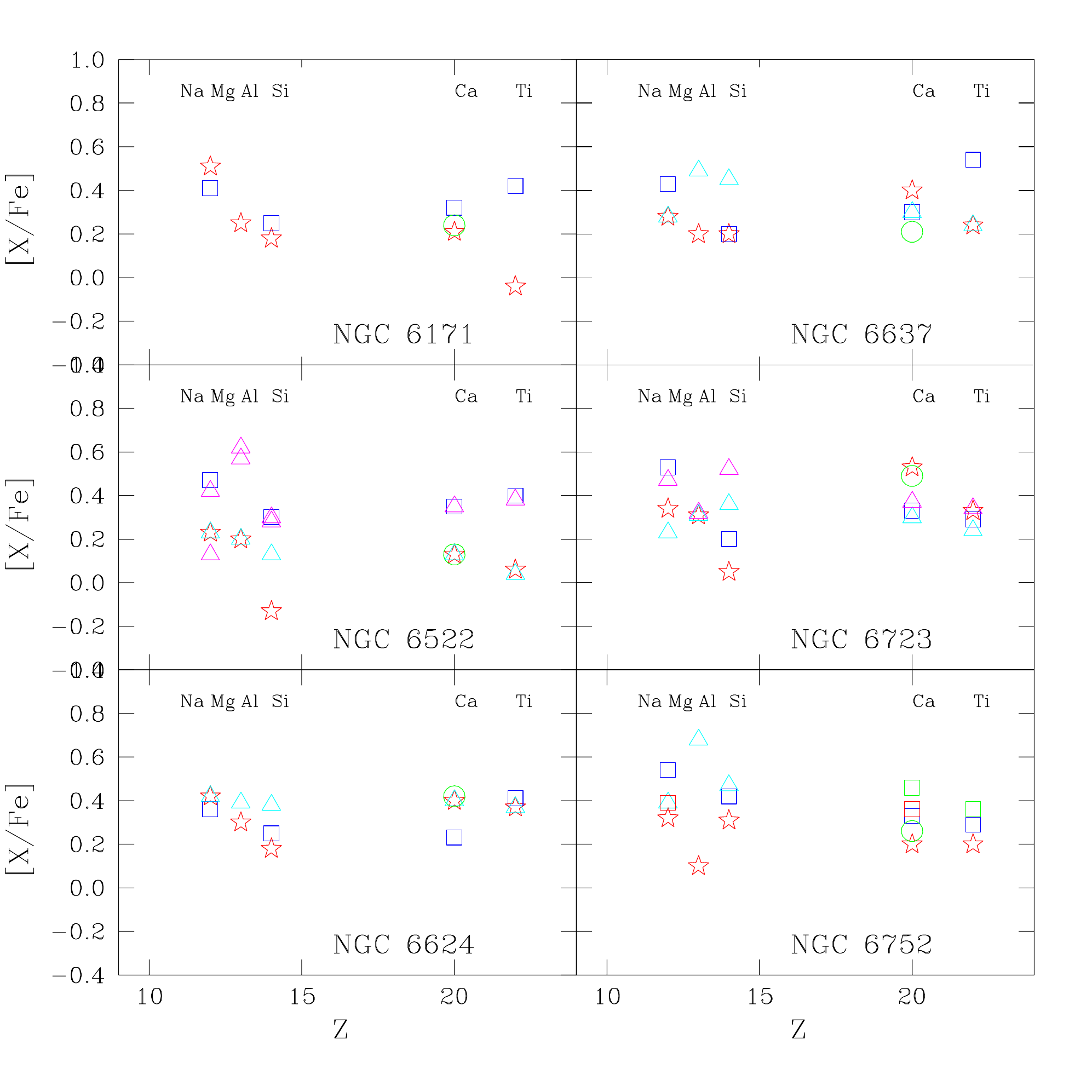}
    \caption{[X/Fe] vs. element number Z. Symbols given in all panels: present results: red open stars;
Conroy et al. (2018): blue
open squares; Usher et al. (2019): green open circles.
Other symbols: cyan open triangles: Barbuy et al. (2014), Valenti et al.
(2011), Lee et al. (2007), Rojas-Arriagada et al. (2016),
Carretta et al. (2012); magenta open triangles: Ness et al. (2014),
Crestani et al. (2019); green open squares: Colucci et al. (2017);
red open squares: Larsen et al. (2017).
}
    \label{fig:pattern}
\end{figure*}

Table \ref{finalabunds} reports the final abundance values
obtained in the present work. In Figure \ref{fig:pattern} these results are compared with
the literature listed in Table \ref{tab2}, for the elements
Mg, Al, Si, Ca and Ti. We do not include Na, for the reasons
explained in Section 5 (i.e. the difficulty in deriving reliable
Na abundances from the lines available), or Ba and Eu because there are very few
literature abundances for these elements for the sample clusters. Figure \ref{fig:pattern}
shows that there is good agreement between the present
results and other abundance derivations from integrated spectra
- stars symbols (ours) compared with squares (others).
A spread of abundances is seen among different high-resolution spectroscopic
analyses: for NGC 6522 the present results are in good agreement with
our previous high-resolution spectroscopic analysis (Barbuy et al. 2014),
and there are differences with Ness et al. (2014). For NGC 6752 the high-resolution spectroscopic analysis by
Carretta et al. (2012), giving abundances for its three stellar populations and
a mean value is plotted in the figure. S. O. Souza et al.
(2019, submitted) found fractions of 25 \%, 43 \%, and 32\% for 1G, 2G, 3G, very similar
to the proportions of 25 \%, 40 \% and 30\% obtained by Milone et al. (2013, and also adopted by Tenorio-Tagle et al. (2019),
therefore a mean of abundances from the three stellar populations appears
suitable. Aluminum has a high abundance, given
their high abundance values for 2G and 3G stars, that we do not confirm
from the integrated spectra.

In conclusion, the derivation of abundances line-by-line from integrated
spectra, for a few suitable lines,
does give results compatible with high-resolution spectroscopic analyses,
and with other integrated spectra analyses.

\begin{table*}
  \caption{Adopted Metallicities and Resulting Mean Abundances for the Elements: Mg, Al, Si, Ca, Ti, Ba, Eu. Na is given in Table \ref{tabna}.}
    \centering
    \setlength{\tabcolsep}{10pt} 
    \label{finalabunds}
    \begin{tabular}{cccccccccccccccccc}
    \tableline
    \hline
    Cluster & \hbox{[Fe/H]} & \hbox{[Mg/Fe]} &
    \hbox{[Al/Fe]} & \hbox{[Si/Fe]} & \hbox{[Ca/Fe]}
& \hbox{[Ti/Fe]}  & \hbox{[Ba/Fe]}    & \hbox{[Eu/Fe]} \\
             \hline
     6171 & -1.02 &  0.51  &  0.25 & 0.18   & 0.21 & -0.04   & 0.00 & 0.50     \\
     6522 & -0.95 &  0.23  &  0.20 & -013   & 0.13 &  0.06   & 0.02 & 0.55     \\
     6624 & -0.69 &  0.42  &  0.30 & 0.18   & 0.40 &  0.37   & 0.10 & 0.00        \\
     6637 & -0.77 &  0.28  &  0.20 & 0.20   & 0.40 &  0.24   & 0.22 & 0.45     \\
     6723 & -0.98 &  0.34  &  0.31 & 0.05   & 0.53 &  0.33   & 0.22 & 0.00     \\
     6752 & -1.53 &  0.32  &  0.10 & 0.31   & 0.20 &  0.20   & 0.00 & 0.50     \\
     \hline
    \end{tabular}
   \label{tab:new_ab}
\end{table*}

\section{Effects of Multiple stellar populations\label{MPs}}

The identification of multiple stellar populations in GCs has been
 recently definitively demonstrated by the so-called chromosome map method, with
 UV photometric data provided in Piotto et al. (2015),
 and explained in detail in Milone et al. (2017). For many of the present clusters
 the identification of percentages of first and second stellar populations are
 available in Milone et al. (2017).
 The reason for the chromosome map to feasible is due to the
  variations in the abundances of C, N, O, as clearly illustrated in Fig. 1 by Piotto et al. (2015).

 The presence of multiple stellar populations together with variations in
 abundances of C, N, O, Na, Al and Mg was identified and described in
 Carretta et al. (2009), Gratton et al. (2012), Renzini et al. (2015), and 
Bastian \& Lardo (2018),
 among others. Anticorrelations Na-O and Mg-Al are observed, where in particular the
 Na-O behavior has been extensively documented (Carretta 2019, Carretta et al. 2009).
Note that for C, N, O, other effects such as stellar evolution along the RGB also affect their abundances.

 \paragraph{Na overabundances}

 The main effect in abundances of second generation stars is 
the Na-O anticorrelation 
as shown in Carretta et al. (2009) and Carretta (2019) for a sample
of GCs. In second generation stars,
 Na is enhanced by amounts between 0.2 and 0.7 (e.g. Campbell et al.
 2013), whereas oxygen is depleted. 
 The proportions of first generation stars relative to total N$_1$/N$_{\rm total}$ were measured by
Milone et al. (2017), and confirmed in R. A. P. Oliveira et al. (2019, in preparation), in a first approximation,
as given in Table \ref{tab1}. If only bonafide members of first and second generations
are taken into account, the proportions can change in some cases (R. A. P. Oliveira et al. 2019, in preparation).
 For the case of NGC 6522, according to Kerber et al. (2018), 84\% of the 
stellar population
 belongs to a second generation, therefore its
first generation fraction should be around N$_1$/N$_{\rm total}$=0.160.
In most cases the second generation is dominant, and this should be seen in
the integrated spectra.

\begin{figure*}[htb!]
     \centering
{\includegraphics[width=0.7\textwidth]{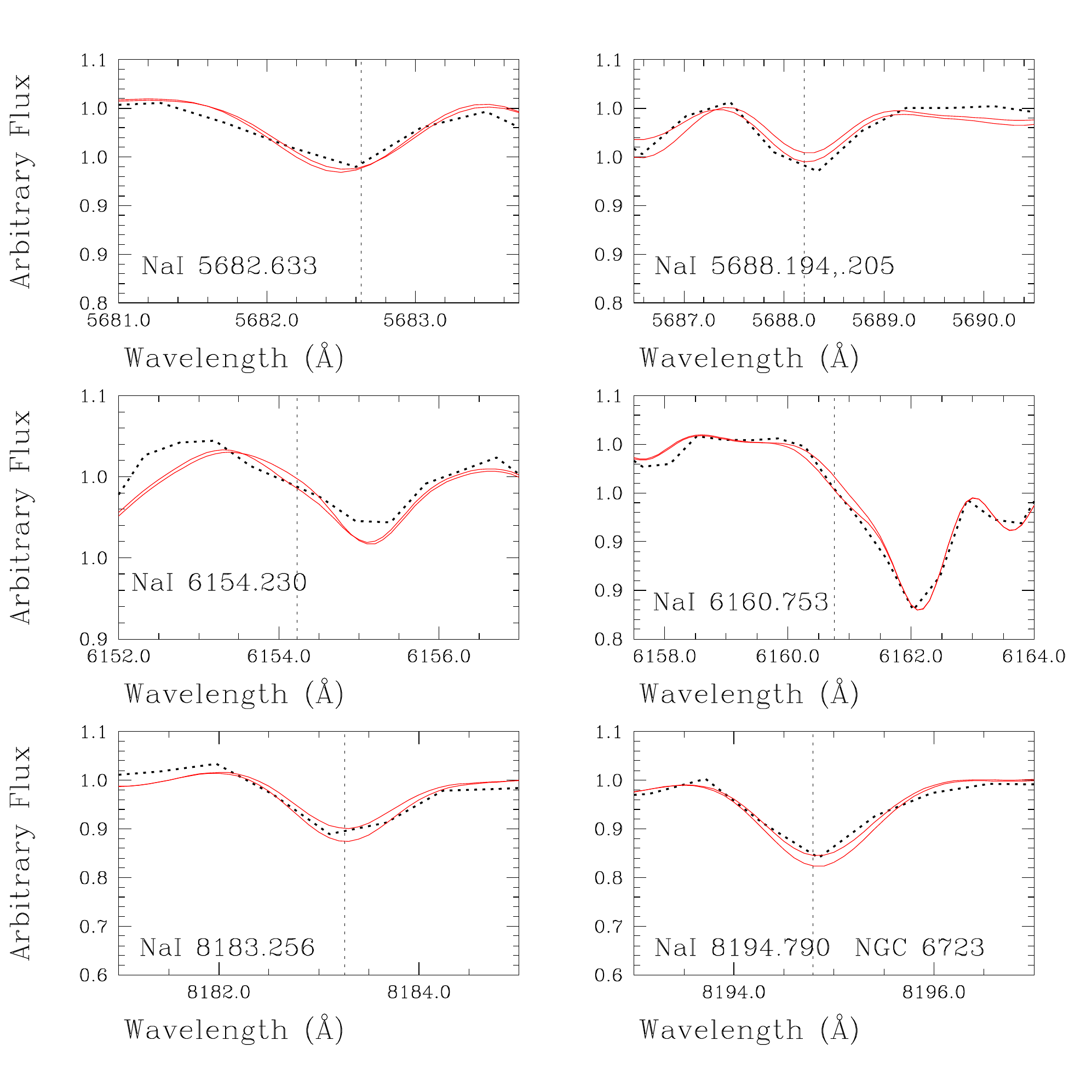}}
\caption{The observed and synthetic spectra for NGC 6723 of Na I lines:
  (a) 5682 {\rm \AA} and 5688 {\rm \AA}; (b) 6154 and 6160 {\rm \AA} and
  (c) 8183 and 8194 {\rm \AA}, computed with [Na/Fe]=0.0 and 0.35.}
     \label{na6723}
 \end{figure*}

Figure \ref{na6723} shows the fit of three doublets of Na I, assuming a mean
[Na/Fe]=0.0 and +0.35, for lines in NGC 6723.
These values are chosen from Table \ref{tab2} where the highest mean
values of [Na/Fe]=0.35 are given for some of the sample clusters.
This Figure for NGC 6723, exemplifies the difficulty in establishing
which is the mean Na abundance. The line 
Na I 5682.63 {\rm \AA} is inconclusive,  line 8194.79 {\rm \AA} indicates
[Na/Fe] = 0.0, whereas 5688.20, 6154.23, 6160.75 {\rm \AA}  indicate [Na/Fe]=+0.35,
and 8183.26 {\rm \AA} is in between. In Table \ref{tabna} we report
the indication of mean Na abundance from this set of lines for the sample
clusters.  The conclusion is that the present data do not allow to draw firm
evidence of second generation stars signatures, and that higher
S/N and higher resolution data would probably allow such a procedure.

Other elements that could be indicators of multiple stellar populations,
Mg and Al, are normally enhanced as found in bulge stars (McWilliam 2016;
Barbuy et al. 2018).
It would be possible to infer multiple stellar population evidence
for a cluster or galaxy if Na is very high and/or Mg is very low.

\begin{table*}
\begin{center}
\caption{Na Abundance from the Main Na I Lines in the Integrated Spectra}
\label{tabna}
\setlength{\tabcolsep}{10pt} 
\begin{tabular}{lcccccccc}
\tableline
\hline
Cluster & 5682.63 {\rm \AA} & 5688.20 {\rm \AA} & 6154.23 {\rm \AA}
 & 6160.75 {\rm \AA} & 8183.25 {\rm \AA} &  8194.79 {\rm \AA} & \\
\hline
NGC 6171 & 0.00   & 0.35  & 0.35 & 0.00  & 0.35 & 0.20 &    \\
NGC 6522 & 0.00   & 0.35  & 0.20 & 0.00  & 0.20 & 0.20 &   \\
NGC 6624 & 0.20   & 0.35  & 0.35 & 0.35  & 0.00 & 0.20 &   \\
NGC 6637 & 0.00   & 0.35 & 0.35  & 0.00  & 0.00 & 0.00 &   \\
NGC 6723 & ---    & 0.35 & 0.35 & 0.35   & 0.20 &  0.00 &   \\
NGC 6752 & 0.00   & 0.35 & --- & 0.35   & 0.20 &  0.35 &   \\
\tableline
\end{tabular}
\end{center}
\end{table*}

\paragraph{IMF and enhancements of Na and TiO features}

 Note that Na enhancements are also used to derive evidence
 on bottom-heavy IMFs (e.g. Conroy \& van Dokkum
 2012). As concerns Na,
given that in our Galactic bulge there is no evidence
 of IMF variations (Wegg et al. 2017, Barbuy et al. 2018a),
 we assume standard IMF throughout this paper.
We believe that Na lines are probably not suited
for IMF deduction, given the effects not only of 
multiple stellar populations discussed in the previous subsection,
but also because Na can be enhanced in the central parts of
galaxies, as it apparently is in our Galactic bulge
(Lecureur et al. 2007, Barbuy et al. 2018a).

Studies on IMF variations are deduced, besides studies using 
Na lines, also from TiO and Wing-Ford FeH bands
(La Barbera et al. 2016).
The FeH bands are located around 1 $\mu$m
(Schiavon et al. 1997), therefore out of the range of the present data.
TiO bands can be investigated with the present method.
We tried to compute TiO bands at 7000-7500 {\rm \AA}, as
previously exemplified in Milone et al. (1995),
applied to the most metal-rich cluster in our sample, NGC 6624.
However, at this metallicity the bands are not strong enough to be
inspected.

\paragraph{Helium abundances}

We have adopted isochrones with primordial helium abundances,
that in the DSED isochrones correspond to Y = 0.247. Despite detection of a small helium enhancement in second generation star for these clusters by Lagioia et al. (2018) and Milone et al. (2018, 2019), the effects on the CMDs, for our purposes, are negligible. The He enhancement in 2G stars of the sample clusters, from Milone et al. (2019), are reported in Table 1. The enhancements are small, therefore the isochrones for 2G, if He enhancement was to be taken into account, would be very close to the 1G enhancement. 
Essentially, if enhanced He is assumed, there is a small decrease in age.
S. O. Souza et al. (2019, submitted) have studied the three stellar populations in
NGC 6752, assuming Y=0.247, 0.257 and 0.288, established by
Milone et al. (2019), and found ages of 13.4, 13.2 and 13.0 for 1G, 2G, and 3G, respectively.

Therefore the effects of helium enhancements can be treated in terms of younger ages, as described in the next section.

\paragraph{Effects of age\label{delta-age}}

\begin{figure*}[ht!]
     \centering
{\includegraphics[width=0.75\textwidth]{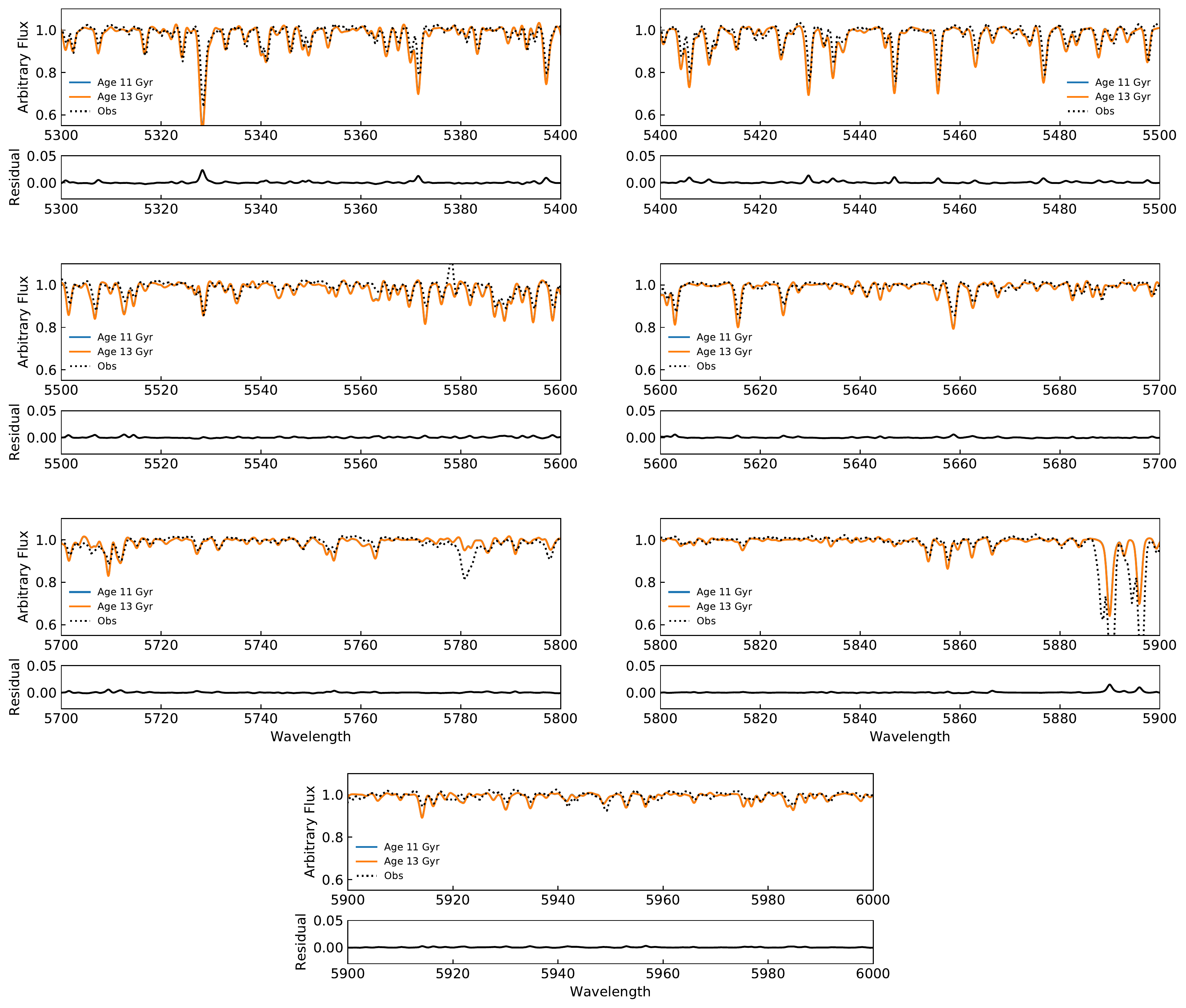}}
\caption{NGC 6522: synthetic spectra for ages of 11 Gyr and 13 Gyr
  in the region $\lambda\lambda$ 5300-6000 {\rm \AA},
 and respective residuals in the lower panels.}
     \label{fig:deltaage}
 \end{figure*}

It is well-known that the effect of age in a CMD 
  is the magnitude of the turn-off point, which is brighter for younger
  ages. In the integrated spectra there is little change in the lines,
  given that the RGB stars dominate the flux. This is
  shown in Figure \ref{fig:deltaage}, with
  the comparison of spectra between 11 Gyr and 13 Gyr for
  NGC~6522, together with the residuals.

\section{Conclusions}

We have built the code SynSSP that combines isochrones for a given
age, metallicity, and [$\alpha$/Fe], to calculations of a series of
synthetic spectra. We are able to reproduce the integrated spectra
of very old GCs in the metallicity range
$-$1.6 $<$ [Fe/H] $<$ $-$0.7.

The main aim of this work is to be able to reproduce the integrated spectra
of well-known GCs, in order to be able to use this
method of building integrated spectra for the analysis of
extragalactic  GCs, such
as developed by e.g., Sakari et al. (2014).

From the present analysis, we can conclude that:

\begin{itemize}

\item It is possible to reproduce
prominent lines of Na, Mg, Al, Si, Ca, Ti, Ba, and Eu,
to derive their abundances from the integrated spectra,
which are in good agreement with literature previous results.
This opens the possibility to apply this method to distant
star clusters in other galaxies.

\item With the abundances of the most important elements being derived,
  it is expected that the residual difference between observed and synthetic
  integrated spectra to be reduced.

\item The effect of age is small among old star clusters. For younger
  star clusters, they would be distinguishable due to their bluer
  colors, and by the presence of other sets of lines.

\item The effect of helium enhancement in second (and subsequent)
  stellar generations is a somewhat younger age, therefore this effect
  can be neglected.
  
\end{itemize}

 Finally, a further
 objective of this work for the future is to use a combination
 of integrated spectra
of SSPs well-tested with our method,
in the building of spectra of galaxies.

\acknowledgments
We acknowledge the anonymous referee for the detailed review and for the helpful suggestions, which allowed us to improve the manuscript. T.C.M acknowledges the post-doctoral FAPESP fellowhip n$^{\circ}$ 2018/03480-7.MT thanks the support of CNPq, process number 307675/2018-1.BB acknowledges FAPESP project 2014/18100-4 and partial financial support from CAPES Financial project 001 and CNPq. SR  acknowledge financial support from FAPESP, process number 2015/50374-0 and partial financial support from CAPES Financial project 001 and CNPq.

\clearpage

\onecolumngrid
\appendix
%

\section{Synthetic Spectra - NGC 6522}

 The full spectral synthesis performed for NGC 6522 in the spectral range of 4500 tp 9000 {\rm \AA} (red lines) are compared to the observed from the WAGGS catalog (Figures 15-19; Usher et al. 2017).
    
\begin{figure*}[ht!]
    \centering
    \includegraphics[width=0.95\textwidth]{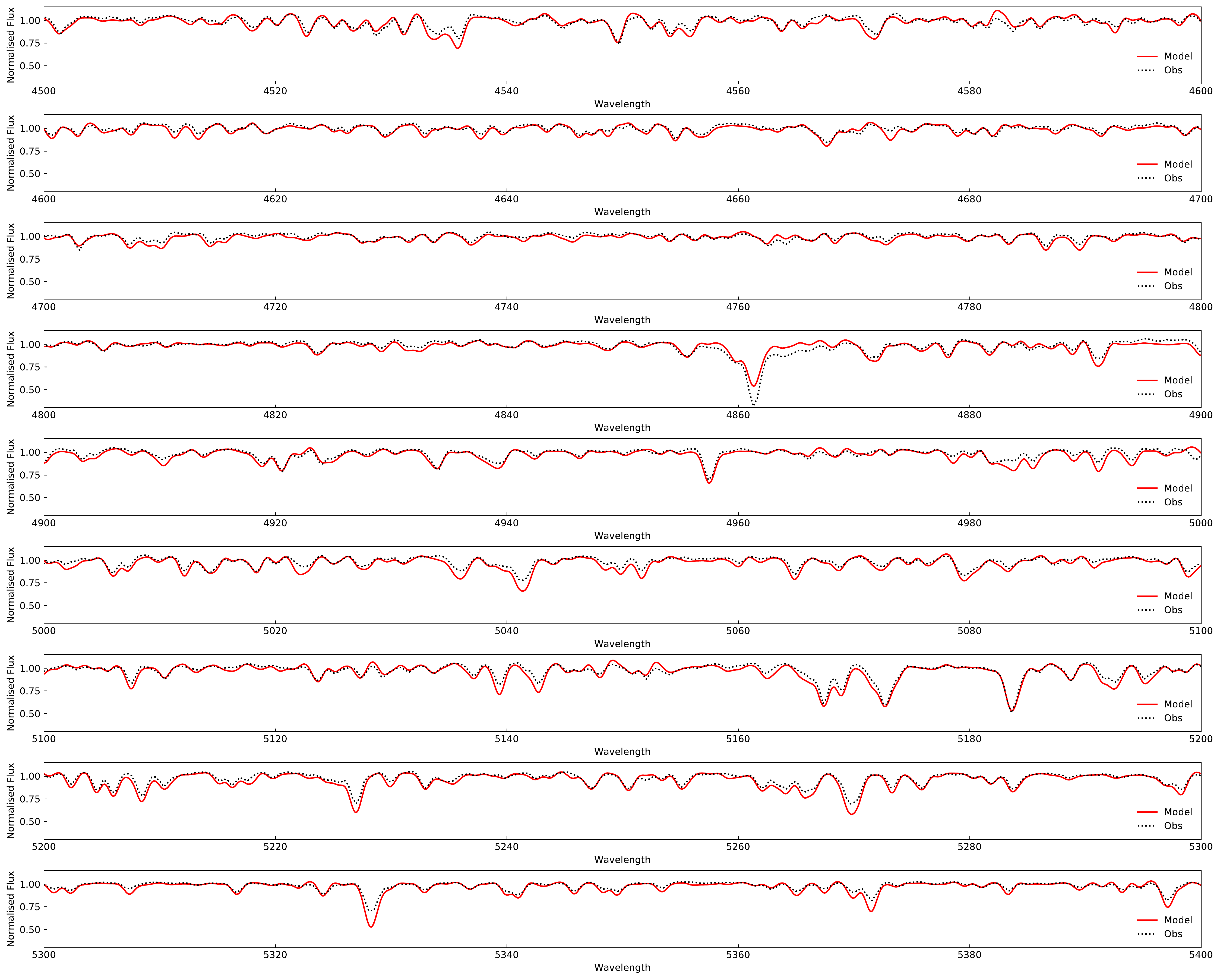}
    \caption{Observed and synthetic spectra for the cluster NGC 6522 in the range  4500 - 5400 {\rm \AA}. Main absorption lines are highlighted. }
    \label{fig:6522_pt1}
\end{figure*}

\clearpage
\begin{figure*}[ht!]
    \centering
    \includegraphics[width=0.95\textwidth]{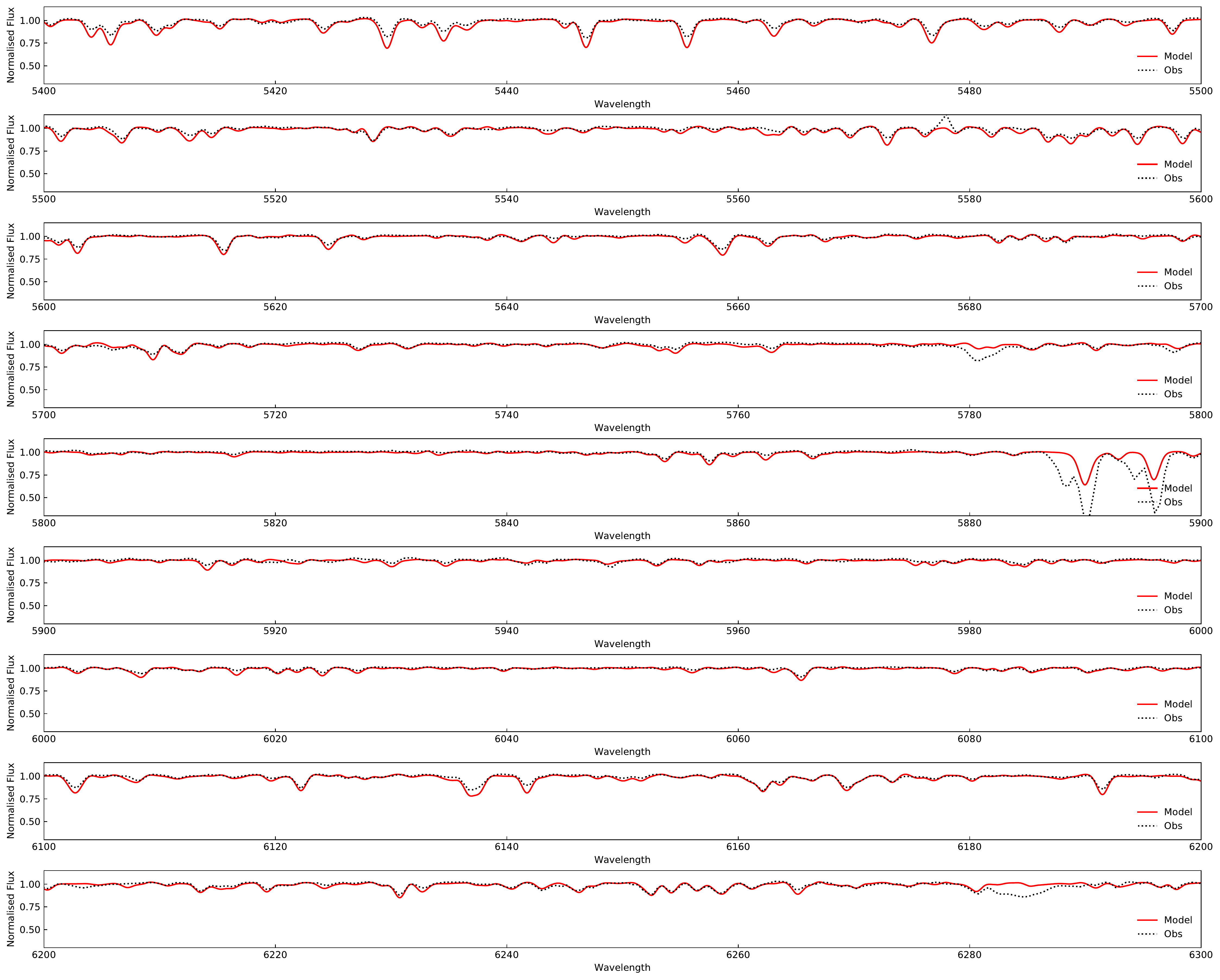}
    \caption{Continuation of Figure \ref{fig:6522_pt1} for spectra in the range  5400-6300 {\rm \AA}. The gray areas represent the region of the telluric line.}
    \label{fig:6522_pt2}
\end{figure*}

\clearpage

\begin{figure*}
    \centering
    \includegraphics[width=0.95\textwidth]{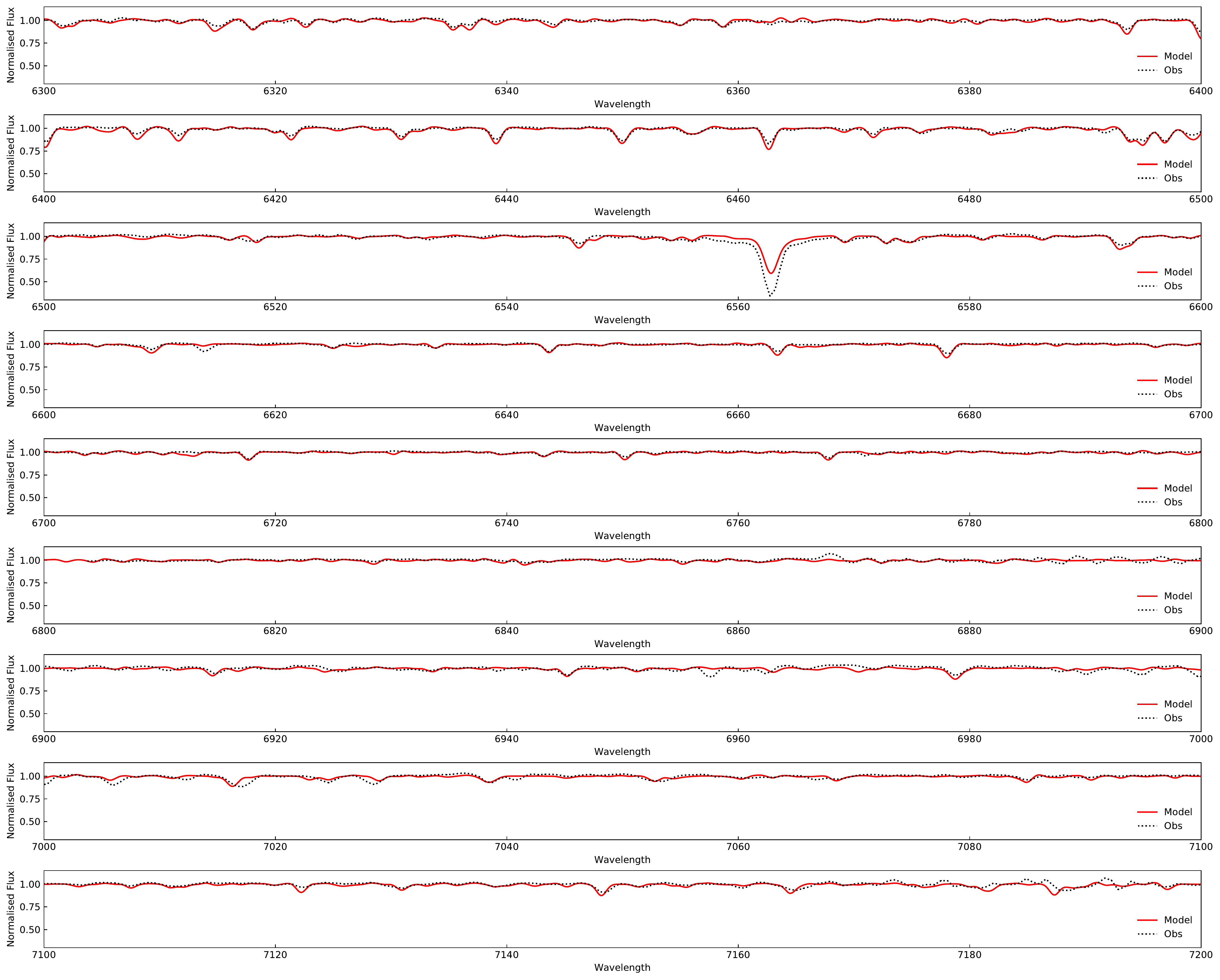}
    \caption{Same as Figure \ref{fig:6522_pt2} for the wavelength range 6300 - 7200 {\rm \AA}.}
    \label{fig:6522_pt3}
\end{figure*}

\clearpage

\begin{figure*}
 \centering
 \includegraphics[width=0.95\textwidth]{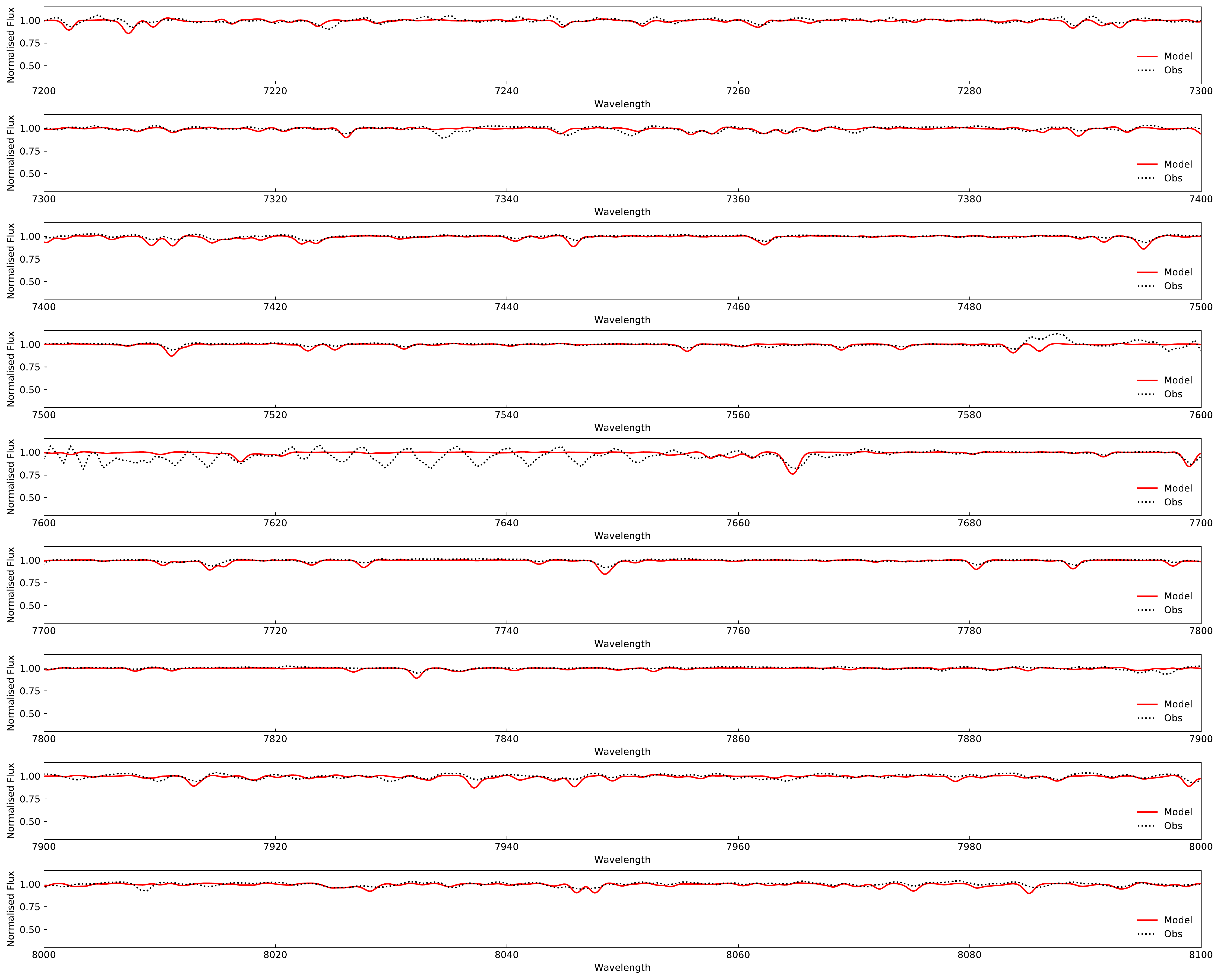}
 \caption{Same as Figure \ref{fig:6522_pt3} for the wavelength range 7200 - 8100 {\rm \AA}.}
 \label{fig:6522_pt4}
\end{figure*}

\clearpage

\begin{figure*}
 \centering
 \includegraphics[width=0.95\textwidth]{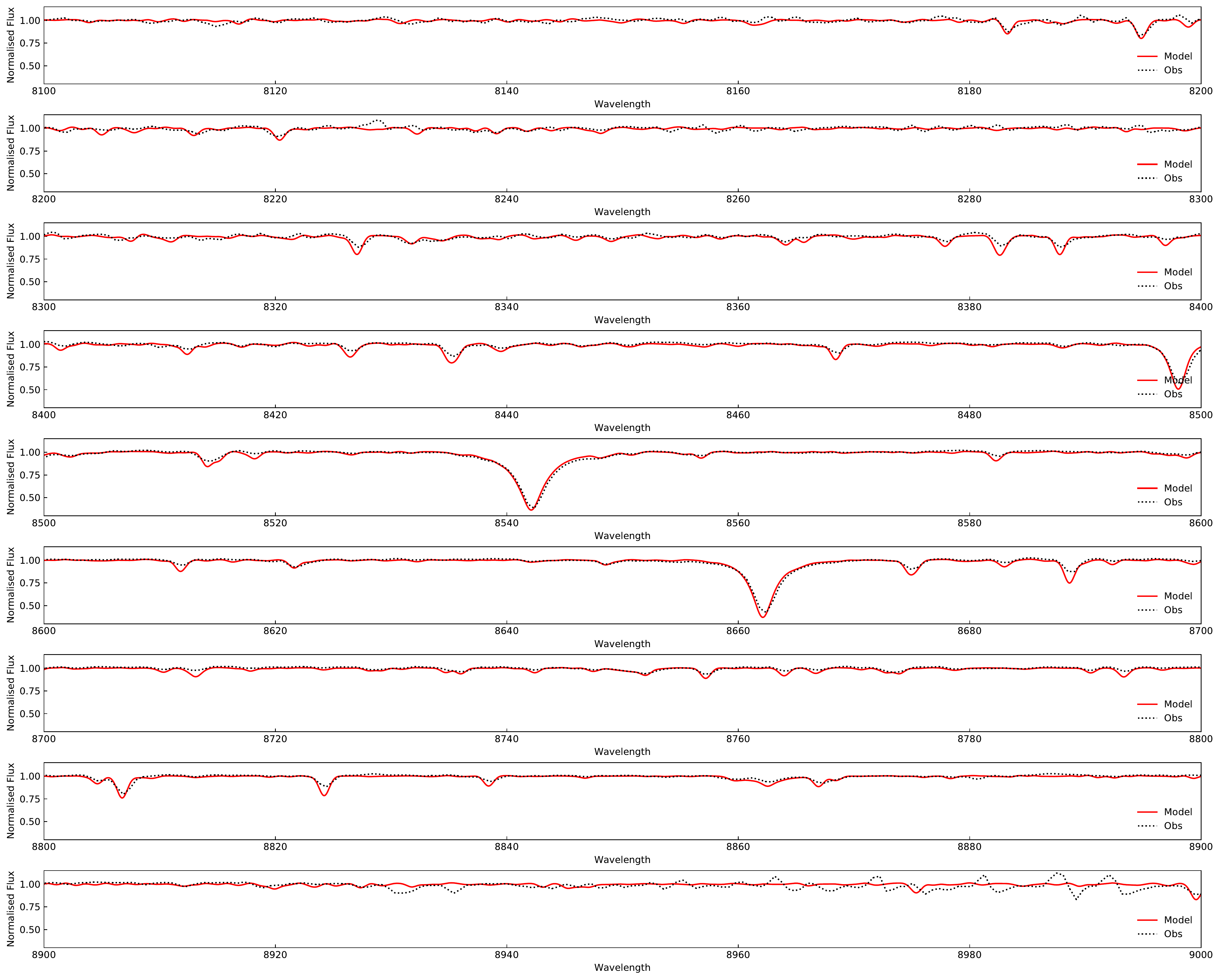}
 \caption{Synthetic and observed spectra for NGC 6522 in the wavelength
   range 8100 - 9000 {\rm \AA}.}
\end{figure*}

\clearpage


\newpage

\end{document}